\documentclass{aa}

\usepackage{graphicx}
\usepackage[caption = false]{subfig}
\usepackage{booktabs}
\usepackage{dcolumn}
\usepackage{txfonts}
\usepackage{xhfill}
\usepackage[titletoc,title]{appendix}
\usepackage{amsmath, amsfonts, amssymb}
\graphicspath{./plots}

\providecommand{\lya}[0]{Ly$\alpha$ }
\providecommand{\lam}[0]{\lambda}
\providecommand{\e}[1]{\times 10^{#1}}
\providecommand{\mc}[1]{\multicolumn{1}{c}{#1}}
\newcommand{\ang}{\mbox{\normalfont\AA }}
\newcommand{\ditto}[1][.4pt]{\xrfill{#1}~~\textquotedbl~~\xrfill{#1}}
\newcommand{\pkg}[1]{\textup{\textsc{#1}}}

\begin{document}
\title{MUSE unravels the ionisation and origin of metal enriched absorbers in the gas halo of a z = 2.92 radio galaxy } 

\author{S. Kolwa \inst{1}
\and J. Vernet \inst{1}
\and C. De Breuck \inst{1}
\and M. Villar-Martin \inst{2,3}
\and A. Humphrey \inst{4}
\and F. Arrigoni-Battaia \inst{5}
\and B. Gullberg \inst{6}
\and T. Falkendal \inst{1,7}
\and G. Drouart \inst{8}
\and M. Lehnert \inst{7}
\and D. Wylezalek \inst{1}
\and A. Man \inst{9}
}     	 

\institute{European Southern Observatory, Karl-Schwarzschild-Str. 2 85748 Garching bei M\"unchen, Germany
\and 
Centro de Astrobiologia (CSIC-INTA), Carretera de Ajalvir, km 4, E-28850 Torrejon de Ardoz, Madrid, Spain
\and
Astro-UAM, UAM, Unidad Asociada CSIC, Facultad de Ciencias, Campus de Cantoblanco, E-28049, Madrid, Spain
\and
Instituto de Astrof\'{I}sica e Ci\^{e}ncias do Espa\c{c}o, Universidade do Porto, CAUP, Rua das Estrelas, Porto, 4150-762, Portugal
\and
Max-Planck-Institut für Astrophysik, Karl-Schwarzschild-Str. 1, 85741
Garching bei München, Germany
\and
Centre for Extragalactic Astronomy, Department of Physics, Durham
University, South Road, Durham, DH1 3LE, UK
\and 
Institut d'Astrophysique de Paris, UMR 7095, CNRS, Universit\'e Pierre et Marie Curie, 98bis boulevard Arago, 75014 Paris, France
\and 
International Centre for Radio Astronomy Research, Curtin University, 1 Turner Avenue, Bentley, Western Australia 6102, Australia
\and 
Dunlap Institute for Astronomy \& Astrophysics, 50 St. George Street, Toronto, ON M5S 3H4, Canada
}

\date{}
\abstract{We have used the Multi-Unit Spectroscopic Explorer (MUSE) to study the circumgalactic medium (CGM) of a z = 2.92 radio galaxy, MRC 0943-242 by parametrising its emitting and absorbing gas. In both \lya $\lam$1216 and \ion{He}{II} $\lam$1640 lines, we observe emission with velocity shifts of $\Delta \varv \simeq-1000$ km s$^{-1}$ from the systemic redshift of the galaxy. These blueshifted components represent kinematically perturbed gas that is aligned with the radio axis, which we interpret as jet-driven outflows. Three of the four known Ly$\alpha$ absorbers are detected at the same velocity as \ion{C}{IV} $\lam\lam1548,1551$ and \ion{N}{V} $\lam\lam1239,1243$ absorbers, proving that the gas is metal enriched more so than previously thought. At the velocity of a strong Ly$\alpha$ absorber with an \ion{H}{I} column of $N_\ion{H}{I}/{\rm cm}^{-2} = 10^{19.2}$ and velocity shift of $\Delta \varv \simeq -400$ km s$^{-1},$ we also detect \ion{Si}{II} $\lam$1260 and \ion{Si}{II} $\lam$1527 absorption, which suggests that the absorbing gas is ionisation bounded. With the added sensitivity of this MUSE observation, we are more capable of adding constraints to absorber column densities and consequently determining what powers their ionisation. To do this, we obtain photoionisation grid models in \pkg{cloudy} which show that AGN radiation is capable of ionising the gas and producing the observed column densities in a gas of metallicity of Z/Z$_\odot \simeq$ 0.01 with a nitrogen abundance a factor of 10 greater than that of hydrogen. This metal-enriched absorbing gas, which is also spatially extended over a projected distance of $r \gtrsim 60$ kpc, is likely to have undergone chemical enrichment through stellar winds that have swept up metals from the interstellar-medium and deposited them in the outer regions of the galaxy's halo.}

\maketitle

\section{Introduction}

High redshift radio galaxies (HzRGs) host very powerful active galactic nuclei (AGN) and occupy the upper echelons of stellar-mass distributions for galaxies across cosmic time \citep{jarvis2001,debreuck2002a,rocca-volmerange2004,seymour2007}. They are often enshrouded by giant \lya emitting haloes that cover regions extending out to $\geq 100$ kpc in projection \citep[e.g.,][]{baum1988,heckman1991,vanbreugel2006,mccarthy1990b,vanojik1996}. These massive haloes also tend to have filamentary and clumpy sub-structures within them \citep{reuland2003}. 

In the case of some HzRGs, extended low surface brightness haloes are found to have quiescent kinematics with line widths and velocity shifts in the order of a few 100 km s$^{-1}$ \citep{villar-martin2003}. Whilst in the high surface brightness regions of these nebulae, perturbed gas kinematics with line widths and velocity shifts that are $> 1000$ km s$^{-1}$ are frequently seen in the extended emission line region (EELR) \citep[e.g.,][]{mccarthy1996,rottgering1997,villar-martin1999a}. Given the alignment of the radio axis with the turbulent kinematics, this has often been interpreted as evidence for jet-gas interactions \citep[e.g.,][]{humphrey2006,morais2017,nesvadba2017a,nesvadba2017b}. Generally, for these reasons, HzRGs are considered some of the best laboratories for studying ionisation and kinematics of gas as well as the mechanisms that power these processes. 

Among these processes are accretion, ionised gas outflows, chemical enrichment and recycling of metal-enriched material which have either been observed or predicted to occur within the circumgalactic medium (see \citet{tumlinson2017} for a formal review). We can find evidence for this within the halo gas surrounding HzRGs that comprises both the interstellar-(ISM) and the circumgalactic mediums (CGM). The latter is our main focus due to it being the gas interface that bridges the gap between the local ISM of a galaxy and the intergalactic medium (IGM) surrounding it.

CGM gas processes have been observed in the form of accretion of IGM gas along the large-scale filaments, into the halo gas of quasars and HzRGs \citep[e.g.,][]{vernet2017,arrigoni-battaia2018}. They have also been seen as what may possibly be gas being expelled from the ISM in the form of AGN-driven outflows or negative feedback \citep[e.g.,][]{holt2008,reuland2007,nesvadba2008,bischetti2017}. Moreover, numerous detections of diffuse, ionised gas around powerful radio galaxies have also been made in observations of ISM and CGM gas \citep[e.g.,][\citealp{gullberg2016} is G16 from hereon]{tadhunter1989,mccarthy1990a,pentericci1999}. While the infall of recycled gas back into the ISM has been predicted by simulations \citep[e.g.,][]{oppenheimer2008,oppenheimer2010,oppenheimer2018} and observed within the haloes of powerful HzRGs \citep[e.g.,][]{humphrey2007,emonts2018}. 

The CGMs of HzRGs are multi-phase, consisting of ionised, neutral and molecular gas. The ionised gas is often located within the EELR, where it has been heated and ionised by star-formation, jet-driven shocks and the AGN, emitting rest-frame ultraviolet (UV)/optical photons \citep[e.g.,][]{villar-martin1997,debreuck1999,debreuck2000,best2000,vernet2001,binette2003,villar-martin2007,humphrey2008a}. Whereas, molecular gas, which is often considered a tracer for star-formation, is detected at mm/sub-mm wavelengths \citep{emonts2015}. The neutral hydrogen component of the CGM can be parametrised by tracing \lya emission and absorption when \ion{H}{I} cannot be directly detected via the 21 cm line \citep[e.g.,][]{barnes2014}.

In the \lya lines detected within HzRGs haloes, the absorption line spectrums are superimposed onto the often bright \lya emission line profiles \citep[e.g.,][]{rottgering1995}. To quantify the absorption in the gas, standard line fitting routines are invoked. With these, the \ion{H}{I} gas causing resonant scattering/absorption of \lya emission can be parametrised in terms of its kinematics and column densities. Studies using this method have found an anti-correlation between the radio sizes of HzRGs and the measured \ion{H}{I} column densities of the absorbers which tend to be primarily blueshifted relative to the systemic velocity of a source \citep{vanojik1997}, which is also observed in \lya blobs surrounding star-forming galaxies \citep{wilman2005}. Furthermore, \citet{wilman2004} have shown that \lya absorbers in HzRGs generally exist in either one of two forms. They are either weak absorbers with column densities ranging from $N_\ion{H}{I}/{\rm cm}^{-2} \simeq 10^{13} - 10^{15}$  and possibly form part of the \lya Forest or they are strong absorbers with $N_\ion{H}{I}/{\rm cm}^{-2} \gtrsim 10^{18}$ that form behind the bow shocks of radio jets, undergoing continual fragmentation as the jet propagates \citep{krause2002,krause2005}.

Evidence of \lya absorption is seen both in alignment with the radio jets and at larger angles from it, proving that \ion{H}{I} absorbing gas can be very extended, covering almost the entire extended emission line region of an HzRG \citep[][\citealp{silva2018b} is S18 from hereon]{humphrey2008b,swinbank2015,silva2018a}. Such absorbers are thought to be shells of extended gas intercepting radiation from the EELR \citep{binette2000}. In these gas shells, \ion{C}{IV} absorption has also been detected, indicating that they have been metal enriched. Often, \ion{C}{IV} column densities are found to be similar in magnitude to those of weak \lya absorbers \citep{villar-martin1999b,jarvis2003,wilman2004}. In addition to being enriched with metals, results from spectro-polarimetry have suggested that at least some of these type of absorbers also contain dust \citep{humphrey2013}.

The subject of this work, MRC 0943-242, is an HzRG at z = 2.92 which has a distinct \lya profile featuring four discrete absorption troughs, first revealed by long-slit, high resolution spectroscopy \citep{rottgering1995}. At even higher resolutions, the four discrete \lya absorbers initially detected have been confirmed with evidence for an asymmetric underlying \lya emission profile also being seen (e.g., \citealp{jarvis2003,wilman2004}; S18). Three of the \lya absorbers fall into the class of weak absorption-line gas as defined by \citet{wilman2004} while one of the \lya absorbers has an unusually high \ion{H}{I} column density of $\simeq$ $10^{19}$ cm$^{-2}.$ 

MUSE (Multi-unit Spectroscopic Explorer) observations of the source have provided a spatially resolved view of the variation in the \lya line and shown that the strong \lya absorber in this source is  extended to radial distances of $r \gtrsim 65$ kpc from the nucleus (e.g., G16). Furthermore, S18 showed that degeneracy between \ion{H}{I} column density and Doppler parameter suggests an alternative \ion{H}{I} column density solution of the strongest absorber i.e. $N_\ion{H}{I}/{\rm cm}^{-2} = 10^{15.2}.$ In the same study, the velocity gradient of the said absorber shows evidence for it being in outflow giving credence to the idea that it formed from an early feedback mechanism \citep{binette2000,jarvis2003}. 

These studies show that the high \ion{H}{I} column density absorber, in particular, is a low metallicity (i.e., Z/Z$_\odot \simeq 0.01-0.02$) gas shell that may have been ejected by previous AGN activity. With respect to the ionisation of the absorber, stellar photoionisation has been said to power the strong \lya absorber \citep{binette2006}. However, much of the progress that has made in determining the ionising mechanism for the absorbers has been hampered by the fact that only column densities of \ion{H}{I} and \ion{C}{IV} were available, at the time. As discussed by \citet{binette2006}, constraints from other lines such as \ion{N}{V} are needed to draw stronger conclusions about the source of ionisation and the chemical enrichment history of the gas.

In this work, we place additional constraints on absorption in \ion{C}{IV}, \ion{N}{V} and \ion{Si}{IV}. This is possible with MUSE which has the sensitivity and spatial resolution needed to measure the size, mass and kinematics of both the emitting and absorbing gas around HzRGs (e.g., \citealp{swinbank2015}; G16; S18). Both G16 and S18 used the MUSE commissioning data which had an on-target time of 1-h. The observations used in this work were obtained over a 4-h on-source integration time and thus have higher signal-to-noise detections of the rest-frame UV lines. Hence, we have been able to detect absorption in resonance lines of lower surface brightness than \lya and \ion{C}{IV} which both G16 and S18 have already studied using MUSE data. 

The paper is structured in the following way. We provide an outline of the data acquisition and reduction steps in Section \ref{section:observations}. Section \ref{section:line-fitting} is dedicated to explaining the details behind the line-fitting routine. In Section \ref{section:best-fit-line}, we present the line models for the emitting and absorbing gas components. In Section \ref{section:morphology-absorbers}, we describe the size, shape, mass and give the ionised fraction of the strongest \lya absorber. The column densities of absorbers in MRC 0943-242 are compared to quasars absorbers in Section \ref{section:hzrg-vs-quasar-abs}. We use photoionisation models to assess whether the AGN can ionise the absorbers to match the observed chemical abundance levels in Section \ref{section:photoionisation-modelling}. We provide an interpretation of our results in Section \ref{section:discussion} and summarise the main findings in Section \ref{section:summary}. 

Throughout the paper, we use $\Lambda$CDM results from the Planck 2015 mission i.e. H$_0$ = 67.8 km s$^{-1}$ Mpc$^{-1},$ $\Omega_{\rm M}$ = 0.308 \citep{Planck2016}. At the redshift of the galaxy, z = 2.92, a projected distance of 1\arcsec subtends a distance of 7.95 kpc. 

\section{Observations and Data Reduction}\label{section:observations}
\subsection{MUSE}
MUSE observations were carried out on the Very Large Telescope (VLT) during 2015-12-14 to 2015-12-15 and 2016-01-14 to 2016-01-18 UT. For the radio galaxy studied in this paper, MRC 0943-242, the observations were obtained under the program run 096.B-0752(A) (PI: Vernet). In the extended wide-field mode, MUSE observes over a wavelength range of $\lam = 4650 - 9300$ $\ang$ without the use of adaptive optics (WFM-NOAO-E). The instrument resolving power is $\lam/\Delta \lam = 1700 - 3400$ which corresponds to a spectral resolution of $\Delta \lam$ = $2.82 - 2.74$ $\ang$ or $\Delta \varv$ $\sim$ $180 - 90$ km s$^{-1}$ (ranging from blue to red). MUSE has a spectral binning of 1.25 $\ang$ ${\rm pix}^{-1}$ and field-of-view (FOV) that is 1 $\times$ 1 arcmin$^2$ in size with a spatial sampling of 0.2 $\times$ 0.2 arcsec$^2$ \citep{bacon2012}. 

Observations of the target, MRC 0943-242, were obtained over 8 $\times$ 30-min observing blocks (OBs) amounting to 4-hr of on-source time. The average seeing disc diameter for the run, under clear conditions, is estimated to be $\rm FWHM$ = (0.74 $\pm$ 0.04)\arcsec. We reduced the raw data in \pkg{esorex} using the MUSE Data Reduction Software (MUSE DRS) pipeline, version 1.6.2 \citep{weilbacher2014}. The data were subsequently processed with the standard MUSE reduction recipe with individual OB exposures being combined at the end of the procedure to create the final datacube. 

Sky subtraction of the data was performed using the principal component analysis (PCA) algorithm, Zurich Atmosphere Purge (\pkg{zap}), which was developed for use on MUSE data \citep{Soto2016}. After a coarse sky subtraction is carried out, the PCA removes sky residuals that result from spaxel to spaxel (spatial pixel) variations in the line spread function.

The MUSE astrometry loses precision due to instrument effects hence we add a slight correction to the astrometry of the final datacube. This was done by identifying field stars in the MUSE FOV from the {\it GAIA DR2} (Data Release 2) catalogue \citep{gaia2016,gaia2018} and computing their astrometry offsets from the {\it GAIA DR2} frame. We calculated these offsets for 8 field stars and used the average right-ascension and declination shifts to reset the central pixel co-ordinates in the MUSE cube header which made the MUSE co-ordinate frame more accurate.

\subsection{UVES}
To supplement this study, we have obtained ancillary data from the VLT instrument, the Ultraviolet Echelle Spectrograph (UVES) \citep{dodorico2000,dekker2000}. These observations show the \lya line in the spectrum of MRC 0943-242 (\citealp{jarvis2003,wilman2004}; S18). During this 3.4-h observation, the red arm of the instrument was used and the configuration was set to a central wavelength of 5800 $\ang.$ The widths of the spatial and spectral binning were 0.5\arcsec and $0.05-0.06$ $\ang.$ The observing conditions resulted in an average seeing disc of 0.8\arcsec. The data was reduced using \pkg{iraf} invoking the standard recipe for echelle spectroscopic data outlined in \citet{churchill1995}.

To obtain the archival 1D spectrum used in this work, the slit was positioned at an angle of $74^\circ$ east of north with a width of 1.2\arcsec, length of 5\arcsec that covered all of the emission from the brightest regions of the gas halo. The spectral resolution ranged between $25,000-40,000$ or $12 - 8$ km s$^{-1}$ ranging from blue to red. 

UVES is better able to resolve the most narrow absorption lines because of its very high spectral resolution. Narrow lines are broadened by instruments, such as MUSE, that have moderate spectral resolution. Hence, the UVES data serves the purpose of allowing us to check the validity of the spectral line-fitting that we perform. 

\section{Spectrum Extraction and Line Fitting Method}\label{section:line-fitting}
\subsection{1D Spectrum Extraction}

\begin{figure} 
\centering
\includegraphics[width=\columnwidth]{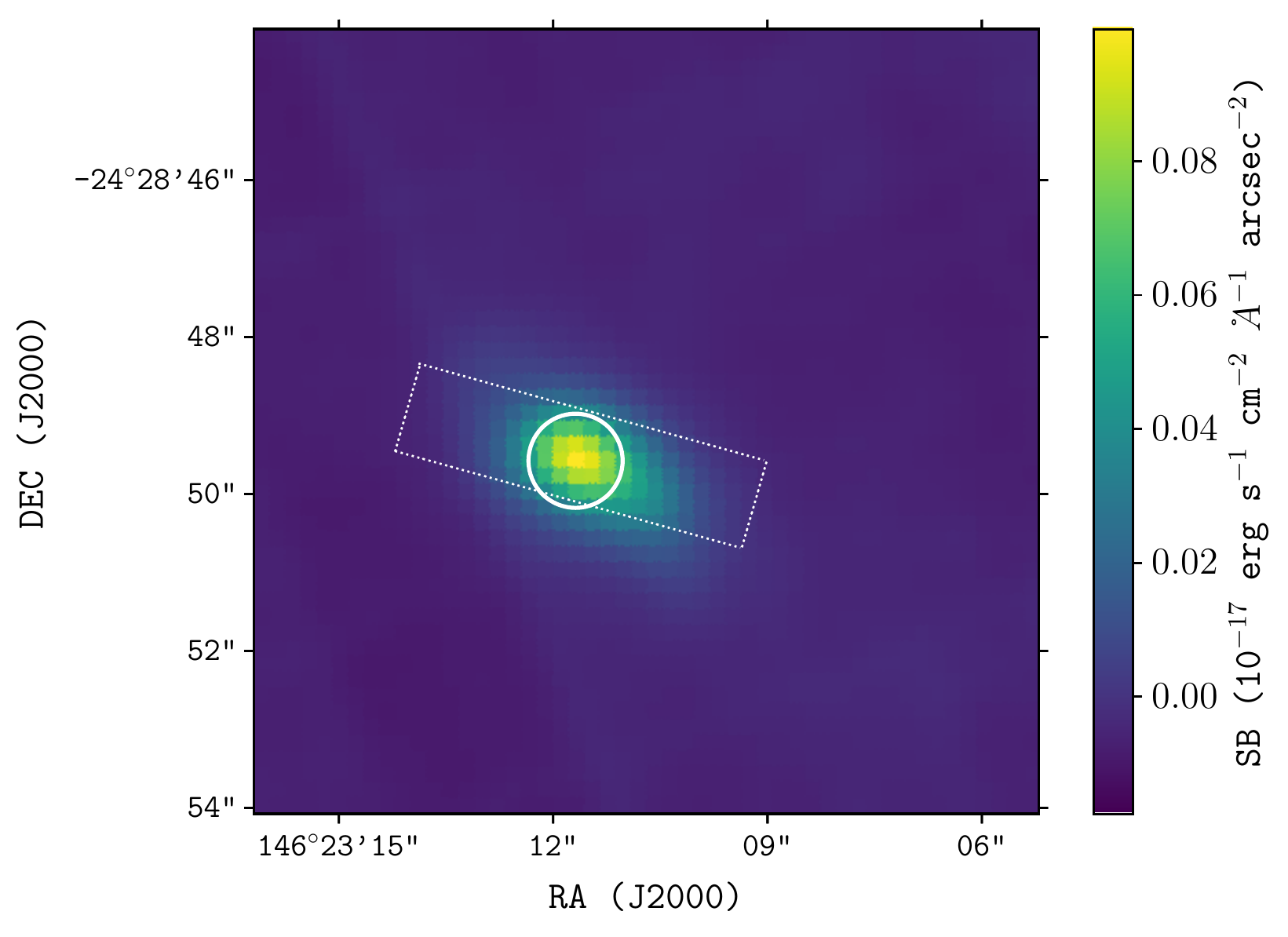}
\caption{MUSE line plus continuum (white light) image of MRC 0943-242. At the centre of the image is the high surface brightness (SB) region of the galaxy halo. The circle represents the aperture from which the spectrum in Fig. \ref{fig:0943-spectrum} is obtained. The UVES slit (dotted outline) has a 1.2\arcsec width and extends over 5\arcsec.}
\label{fig:0943-continuum}
\end{figure}

\begin{figure*} 
\centering
\includegraphics[width=\textwidth]{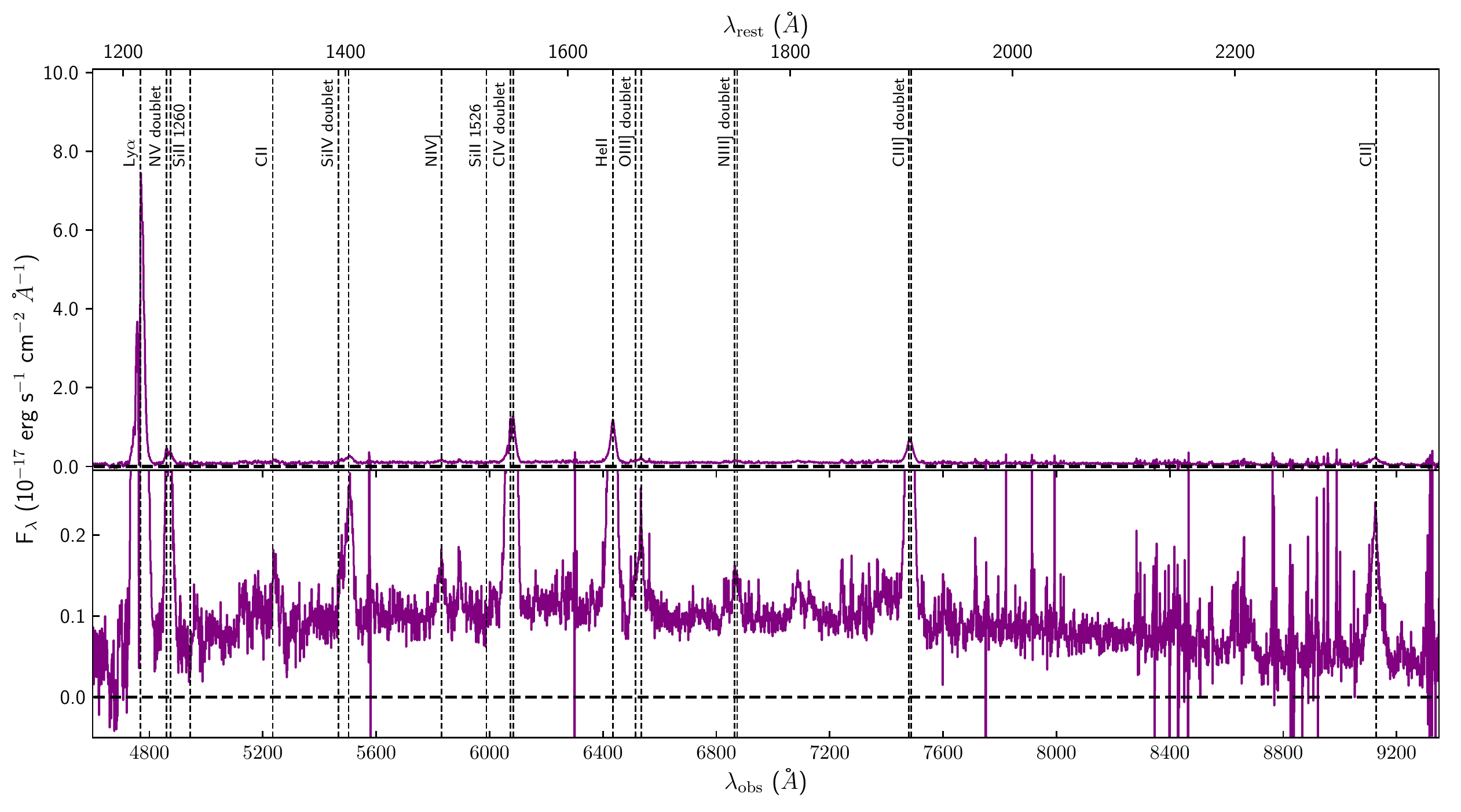}
\caption{A MUSE spectrum of the high surface brightness region in the halo of MRC 0943-243. The spectrum contains several rest-frame UV lines which are labelled and indicated by the dashed vertical lines. The upper panel shows the entire flux density range for the spectrum, while the lower panel covers only the low flux density range which shows the lower S/N lines more clearly.}
\label{fig:0943-spectrum}
\end{figure*}

With the goal of studying the absorbing gas surrounding the host galaxy of MRC 0943-242 (which we refer to as Yggdrasil, following the naming convention provided in G16), we extract a 1D spectrum from a sight-line in the MUSE datacube where the surface brightness of rest-frame UV emission is highest. We refer to it as the high surface brightness region (HSBR), within the gas halo of Yggdrasil, which we are interested in studying (shown in Fig. \ref{fig:0943-continuum}). Although, it sits within the brightest parts of the gas halo, it is not the location of the AGN because this cannot be easily inferred. From this HSBR, we extract the 1D spectrum over an aperture of radius is R=3 spaxels or R=0.6\arcsec centred at the brightest spatial pixel or spaxel located at the co-ordinates, ($\alpha, \delta$) = ($145^\circ 23\arcmin 11.70\arcsec, -24^\circ 28\arcmin 49.58\arcsec$). Collapsing the sub-cube spatially by summing the flux over all spaxels, we obtain the rest-frame UV spectrum shown in Fig. \ref{fig:0943-spectrum}. 

\subsection{Line-fitting Procedure}
The goal of this work is to parametrize the physical conditions of gas in the halo of the radio galaxy, Yggdrasil. To do this, we use the line profiles of ions that undergo resonance transitions i.e. \lya $\lam$1216, \ion{C}{IV} $\lam\lam1548, 1551$, \ion{N}{V} $\lam\lam1238, 1243$ and \ion{Si}{IV} $\lam\lam1393, 1402 $ (see Fig. \ref{fig:0943-spectrum}). We apply line-fitting procedures that combine Gaussian and Voigt models in order to characterise the emission and absorption simultaneously. The emergent emission is $F_\lam = F_{\lam,0} e^{-(\tau_{\lam,1} + ... + \tau_{\lam,n}) }$ for $n$ absorbers, where the unabsorbed emission, $F_{\lam,0},$ is denoted by the Gaussian function, 
\begin{equation}
F_{\lam,0} = \frac{ F }{ \sigma_\lam \sqrt{2\pi}} \exp\left[{-\frac{1}{2}\left(\frac{\lam - \lam_0}{\sigma_\lam}\right)^2}\right]
\end{equation}
where $F$ is the integrated flux of the underlying emission, $\lam_0$ the Gaussian line centre, $\sigma_\lam$ the line width, and $\lam$ the wavelength.
The absorption is quantified by the optical depth, $\tau_\lam,$ which is denoted by the Voigt-Hjerting function,
\begin{equation}
\tau_\lam = \frac{ N\sqrt{\pi} e^2 f \lam_0^2 }{ \Delta\lam_D m_e c^2 }H(a,u),
\end{equation}
where $N$ is the column density, $e$ (electron charge), $m_e$ (electron mass) and $c$ (light speed) are fundamental constants, and $f$ is the oscillator strength. $H(a,u)$ is the Hjerting function in which $a \equiv\frac{\Gamma\lam_0^2}{4\pi c \Delta\lam_D}$ and $u \equiv\frac{(\lam - \lam_0)}{\Delta\lam_D}$ such that $\Gamma$ is the Lorentzian width, $\Delta\lam_D$ is the Doppler parameter (also $b$ parameter), and $\lam - \lam_0,$ is the frequency shift from the line centre ($\lam_0$). $H(a,u)$ is obtained from the \citet{tepper-garcia2006} approximation which is well suited for absorption systems that have column densities $ N/{\rm cm}^{-2} \leq10^{22}.$ When fitting the absorption, we also make the simplifying assumption that each cloud of absorbing gas covers the emission line region with a unity covering factor (C $\simeq$ 1.0).

We use the \pkg{python} package, \pkg{lmfit}, to carry out the non-linear least squares fitting \citep{newville2016}. Our fitting method of choice is the Levenberg-Marquardt algorithm which performs the $\chi^2$-minimization that yields our best fit results. In the figures, we report reduced chi-squared value, $\chi_\nu^2 = \chi^2/(N-N_i),$ where $N$ is the number of data points and $N_i$ is the number of free parameters. For each line, we calculate the local continuum level by masking the line emission and fitting a first-order polynomial to the surrounding continuum. The first order polynomial is subtracted from both the line and continuum of \ion{He}{II} and Ly$\alpha$ which are bright enough that not fitting the continuum ends up having little effect on the overall fit. For \ion{C}{IV}, \ion{N}{V} and  \ion{Si}{IV}, which are lower in surface brightness (than \lya and \ion{He}{II}) the continuum is more important. 

Fitting Gaussian and Voigt functions simultaneously results in the composite model having a high number of fit parameters. To prevent over-fitting and also obtain a physical result, we use \lya and \ion{He}{II} as initial guesses when fitting underlying emission and absorption profiles to the \ion{C}{IV}, \ion{N}{V} and \ion{Si}{IV}+\ion{O}{IV]} lines. 

From the \ion{He}{II} fit (described later in section \ref{section:Heii-fit}), we obtain the best fit results for Gaussian fluxes (F$_\lam$), line widths ($\sigma_\lam$) and line centres ($\lam_0$). $\Delta\lam_{0,\ion{He}{II}}$ is the error on the fitted \ion{He}{II} line centre such that $\lam_{0,\ion{He}{II}} = 6436.09 \pm 0.30$ $\ang.$ Note that the line centres are in agreement with the \ion{He}{II} systemic velocity (or zero velocity which is fixed to the systemic redshift of the galaxy) within its uncertainties i.e. $\Delta\lam_{0,\ion{He}{II}} = 0.30$ $\ang$ = 10.3 km s$^{-1},$ under the assumption that all the emission originates from the centre of the halo. The \ion{He}{II} fit results for line width and the redshift of the line centre are set as initial guesses in Gaussian fits for emission in the resonant lines, \ion{C}{IV}, \ion{N}{V}, \ion{Si}{IV} as well as the non-resonant \ion{O}{IV]} which emits as part of the \ion{Si}{IV}+\ion{O}{IV]} intercombination line. Furthermore, we ensure that $F_\lam$ and $\sigma_\lam$ remain positive. 

To fit the \lya absorption in the MUSE spectrum, we have used the best fit parameters from literature \citep[i.e.,][]{jarvis2003,wilman2004} as initial guesses in the fitting procedure. The best fit \lya absorber redshifts have been passed on as initial guesses for absorber redshifts in the \ion{C}{IV}, \ion{N}{V} and \ion{Si}{IV} line profiles. The initial guesses for column densities and Doppler parameters are based on rough estimates from the literature (i.e., \citealp{jarvis2003,wilman2004}, G16; S18). All three Voigt parameters have a limited parameter space over which a solution can be obtained. These parameter constraints are summarised in Table \ref{table:absorption-limits}. 

In the fitting routine, some absorber redshifts are given more freedom to vary over a given parameter space than others because 
the ionisation energies of the different gas tracers, \ion{H}{I}, \ion{C}{IV}, \ion{N}{V} and \ion{Si}{IV}, differ greatly such that E$_\ion{H}{I}$ = 13.6 eV, E$_\ion{C}{IV}$ = 64.5 eV, E$_\ion{N}{V}$ = 97.9 eV and E$_\ion{Si}{IV}$ = 45.1 eV. It would be unrealistic to expect them to be exist at exactly the same redshifts. 

The minimum Doppler parameter permissible in the line-fitting is set by the lower limit of the MUSE spectral resolution which suggests that an approximate lower limit of $b_{\rm min} \sim \sigma_\lam = 90$ km s$^{-1}$ / 2.3548 $\simeq$ 40 km s$^{-1}.$ We set a conservative upper limit of $b_{\rm max} = 400$ km s$^{-1}$ for \ion{C}{IV}, \ion{N}{V} and \ion{Si}{IV} with the understanding that the absorbing gas is kinematically quiet in relation to perturbed gas regions that have FWHM $\geq$ 1000 km s$^{-1}.$ 

In general, \lya absorbers in HzRG gas nebulae are found to have low column densities of $N_\ion{H}{I}/{\rm cm}^{-2} = 10^{13} - 10^{15}$ or be optically thick with column densities of $N_\ion{H}{I}/{\rm cm}^{-2} > 10^{18}$  \citep{wilman2004}. Since \ion{H}{I} is frequently more abundant than metals in halo gas environments, we can expect the column densities of the metal ions to be lower. Hence, we set the lower and upper limits for permissible fitted column densities as, $N/{\rm cm}^{-2} = 10^{12} - 10^{20}$ cm$^{-2}.$ A full summary of the initial boundary conditions in the fitting procedure is given in Table \ref{table:absorption-limits}. 

\section{Best fit Line Models}\label{section:best-fit-line}
\subsection{\ion{He}{II}}\label{section:Heii-fit}
\ion{He}{II} $\lam$1640 is the brightest non-resonant emission line and forms the basis our estimation of the underlying emission profiles of the lines we study here i.e. Ly$\alpha,$ \ion{C}{IV}, \ion{N}{V} and \ion{Si}{IV}. In the extracted spectrum, we have obtained a detection of \ion{He}{II} emission which forms through cascade recombination of He$^{++},$ which emits a non-resonant photon. The \ion{He}{II} lines is often used to determine the systemic redshift of a galaxy (e.g., \citealp{reuland2007,swinbank2015}; S18). Although \lya is brighter than \ion{He}{II}, we refrain from using \lya for this purpose because of its susceptibility to radiative transfer effects that clearly affect the emergent line emission.

\begin{figure} 
\centering
\includegraphics[width=\columnwidth]{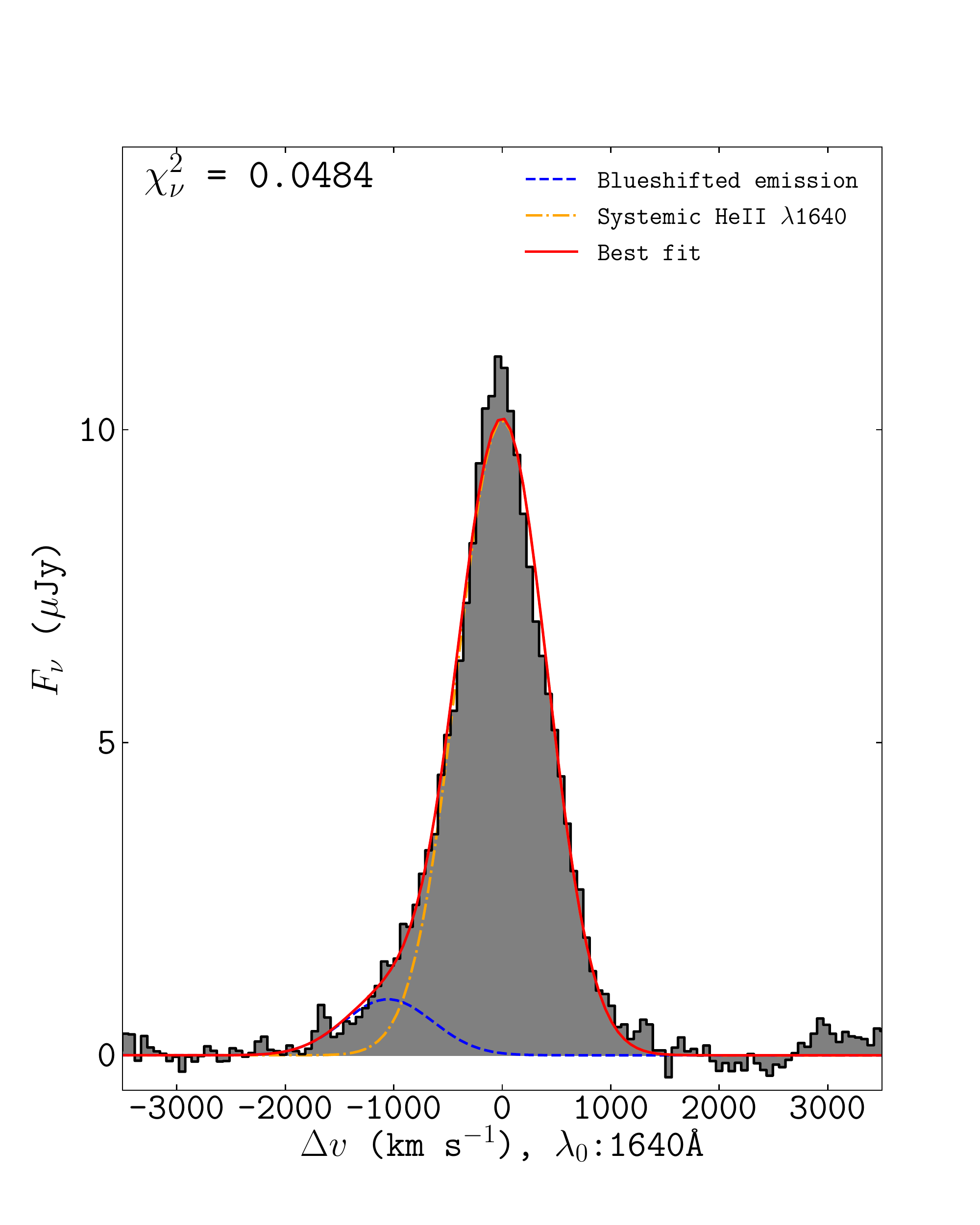}
\caption{The \ion{He}{II} $\lam$1640 line in the MUSE spectrum (shown in Fig. \ref{fig:0943-spectrum}). The line profile has been continuum subtracted. The best fit model (red) consists of two Gaussian profiles. The first Gaussian component (blue) models the excess blue wing emission at the negative velocities, while the second component (orange) models the emission at the systemic velocity of the galaxy. }
\label{fig:HeII-line}
\end{figure}

The \ion{He}{II} line, shown in Fig. \ref{fig:HeII-line}, has an excess of emission at $\Delta \varv \sim-1000$ km s$^{-1}$ which is not present at $\Delta \varv \sim1000$ km s$^{-1}$ indicating asymmetry which has been identified before in \citet{jarvis2003} where the excess was found to have little effect on the final fit. Our 1D spectrum has brighter line detections and thus, to account for the excess emission in the wing, we fit the \ion{He}{II} line with two Gaussian profiles: one for emission blueshifted relative to and another for emission originating from the systemic velocity. This method has been employed frequently for fitting asymmetric line profiles from various gas phases \citep[e.g.,][]{mullaney2013,cicone2014,rakshit2018,hernandez-garcia2018,perna2019}. 

We have estimated the central velocity of the excess emission in \ion{He}{II} by spatially identifying where its emission peaks in surface brightness (explained further in section \ref{method:blueshifted-emission}). The best fit Gaussian parameters obtained from the spatially offset region are used to fit the blueshifted emission as a second component to the emission from the high surface brightness region (HSBR) at the systemic velocity. The result of this fitting procedure is shown in Fig. \ref{fig:HeII-line} where the additional component is broadened to a line width of FWHM $\simeq1000$ km s$^{-1}$ and also blueshifted from the systemic velocity. From the \ion{He}{II} best fit result, we obtain a fiducial systemic redshift of z$_{\rm sys}$ = $2.9235 \pm 0.0001$ that denotes the zero velocity for all the lines identified in Fig. \ref{fig:0943-spectrum}.

\subsection{\ion{H}{I} \lya}\label{section:Lya-fit}

\begin{figure} 
\centering
\includegraphics[width=\columnwidth]{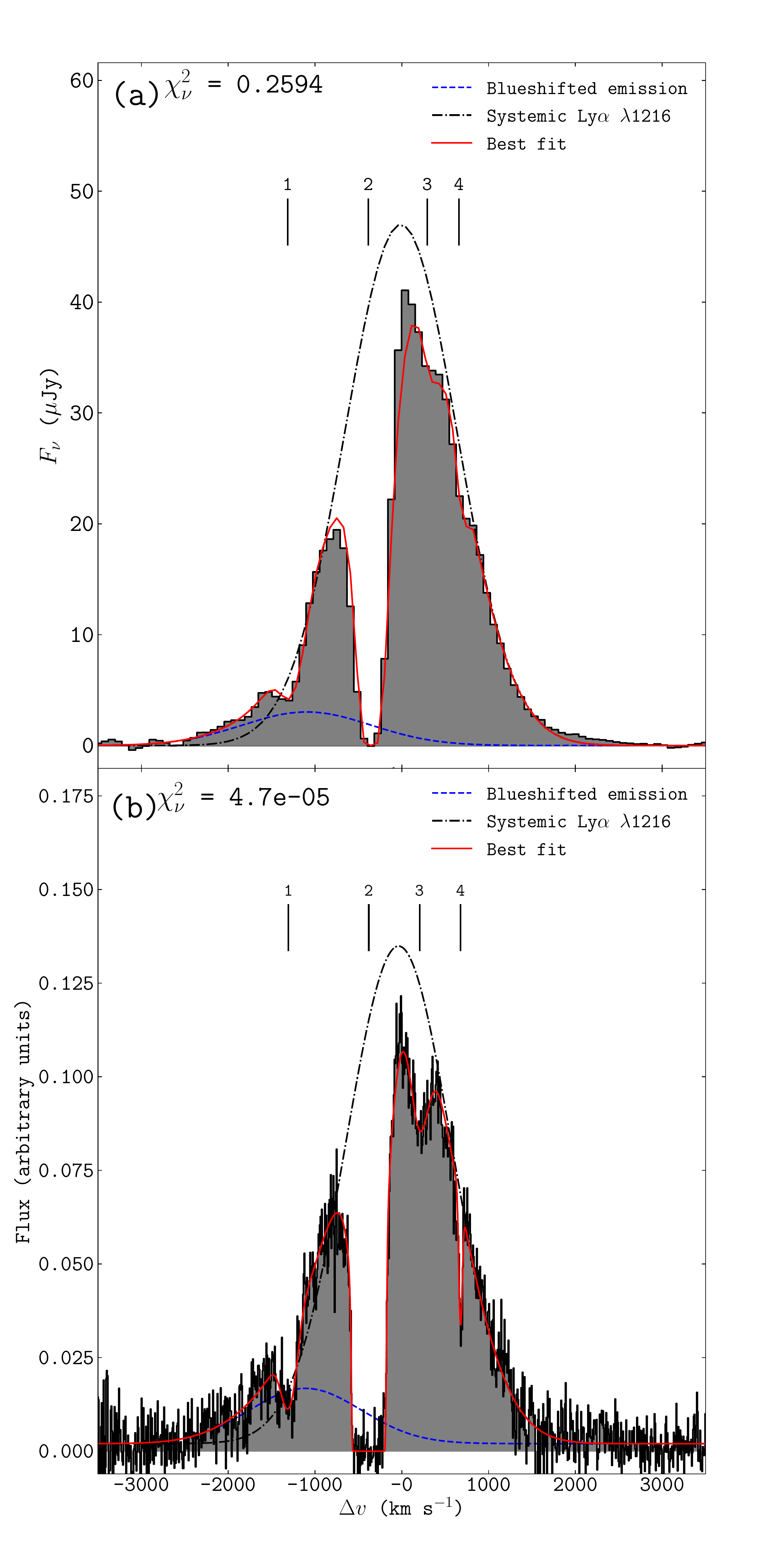}
\caption{MUSE-detected, continuum-subtracted \lya ($upper$) and ancillary UVES-detected \lya ($lower$) are shown. The best fit line model (red) combines \lya emission and consists of blueshifted and systemic components, as well as the four known absorption troughs.}
\label{fig:Lya-line-muse-uves}
\end{figure}

\ion{H}{I} Ly$\alpha$ $\lam$1216, is the brightest line in the rest-frame UV spectrum. We fit the \lya emission envelope with a double Gaussian: one to the non-absorbed singlet emission at the line centre and another to the strong blue wing emission at $\Delta \varv \simeq -1000$ km s$^{-1}$ (see Fig. \ref{fig:Lya-line-muse-uves}). The motivation for including a second emission component to the \lya fit is two-fold: a) it is well detected in the emission line, \ion{He}{II} and therefore likely to also emit in \lya and b) there is an asymmetry between emission at the blue and red wings of Ly$\alpha.$ Once fit, we see that the inclusion of a second Gaussian, improves the \lya emission fit. 

The updated Gaussian fit describes the \lya profile well in both MUSE and UVES spectrums and the best fit parameters are in good agreement with the literature (see Tables \ref{table:absorption-fits-uves} and \ref{table:emission-fits-uves}). This is expected since \lya is affected by radiative transfer effects and therefore less likely to have a symmetric emission profile. There is also an additional underlying kinematic component causing the blue wing excess, based on the \ion{He}{II} line fit (see Fig. \ref{fig:HeII-line}). 

To account for the absorbers, we fit four Voigt profiles, as has been done in previous works on this topic (i.e., \citealp{vanojik1997,jarvis2003,wilman2004}; G16; B18). We also convolve the Voigt profiles with the line-spread function of MUSE, using a Fast-Fourier-Transform similar to that used to convolve Voigt and LSF (Line Spread Function) profiles in the package, \pkg{vpfit} \citep{krogager2018}. The LSF or instrumental profile (IP) of MUSE, when convolved with the Voigt profile, has an average Gaussian width of $<\sigma_\lam>$ = 2.65 $\ang.$ 

We show the line-fitting result for \lya  in Fig. \ref{fig:Lya-line-muse-uves}(a). Below this in Fig. \ref{fig:Lya-line-muse-uves}(b), we show a similar fit to the \lya profile as it was detected with UVES which has an LSF with an average Gaussian width of $<\sigma_\lam>$ = 0.3 $\ang.$ 

The \lya absorber redshifts are used to predict the most probable redshifts for the absorbers, in general. In particular, the redshifts of the absorbers associated with resonance metal ions, \ion{C}{IV}, \ion{N}{V} and \ion{Si}{IV}, are constrained to stay within $4\Delta{\rm z_{sys}}$  (where $\Delta{\rm z_{sys}} =1.35\e{-4}$) agreement of the measured \lya absorber redshifts. This is done following the hypothesis that \lya absorption occurs within roughly the same volume of gas as absorption of photons associated with the resonant transitions. 

For this to occur, there would need to be a very strong ionising continuum to produce all of the observed ions which is possible with ionisation by the AGN. The different ionisation energies also imply that fixing the absorbers to exactly the same redshift is unrealistic. This is why we allow the fitted absorber redshifts to vary over a parameter space that depends on how tightly a parameter needs to be constrained. 

\subsection{\ion{C}{IV} and \ion{N}{V}}\label{section:CIV-NV-fit}

We have obtained detections of \ion{C}{IV} $\lam\lam1548,1551$ and \ion{N}{V} $\lam\lam1238,1243$. Given that \lya absorber 4 is narrow, it is likely to suffer the highest degree of instrumental broadening in MUSE, as seen in Fig. \ref{fig:Lya-line-muse-uves}(a). Hence for \ion{C}{IV} and \ion{N}{V}, we do not include a fourth absorber in the model. Indeed, at the MUSE spectral resolution we expect to lose any absorption signal from such a narrow absorber. The Voigt profiles have, as with Ly$\alpha,$ been convolved with the LSF of MUSE. 

The observed emission in the doublet lines, \ion{C}{IV} and \ion{N}{V}, is fit with two Gaussian functions that are constrained according to atomic physics (all constraints are summarised in Table \ref{table:absorption-rules}). The local continuum has been estimated using a  separate linear polynomial fit (to the continuum only, with line emission masked). The continuum is thus fixed during fitting. Additionally, in \ion{C}{IV}, we fit a second component to each of the doublet lines to account for the blueshifted emission seen in \ion{He}{II} which has a similar S/N level as \ion{C}{IV}. \ion{N}{V} has a lower S/N than both of these lines hence the blueshifted emission is likely to be negligible in this fit. 

We obtain atomic constants such as rest wavelengths and oscillator strengths from the database provided by \citet{cashman2017}. For doublet lines, the ratios of line centres ($\lam_1/\lam_2$) and rest-frame wavelengths are fixed to one another. The doublet emission originates from the same gas hence doublet line widths are equal i.e. $\sigma_{\lam,1} = \sigma_{\lam,2}$. The doublet ratios (DR = $F_1/F_2$) are fixed to the those of the oscillator strengths such that are DR = 2. Doublet ratios of 2 observed frequently in quasar absorption lines where. This is particularly true for \ion{C}{IV} lines whose doublet ratios vary from 2 to 1 as ones goes from the linear to the saturated absorption regimes \citep{peroux2004}. Hence, assuming that \ion{C}{IV} and \ion{N}{V} absorbers 1 and 3 are not saturated and have a unity covering factor, C $\simeq$ 1.0 since their emission does not reach zero flux level at the absorber velocities, we set DR = 2 when fitting.

\begin{figure} 
\centering
\includegraphics[width=\columnwidth]{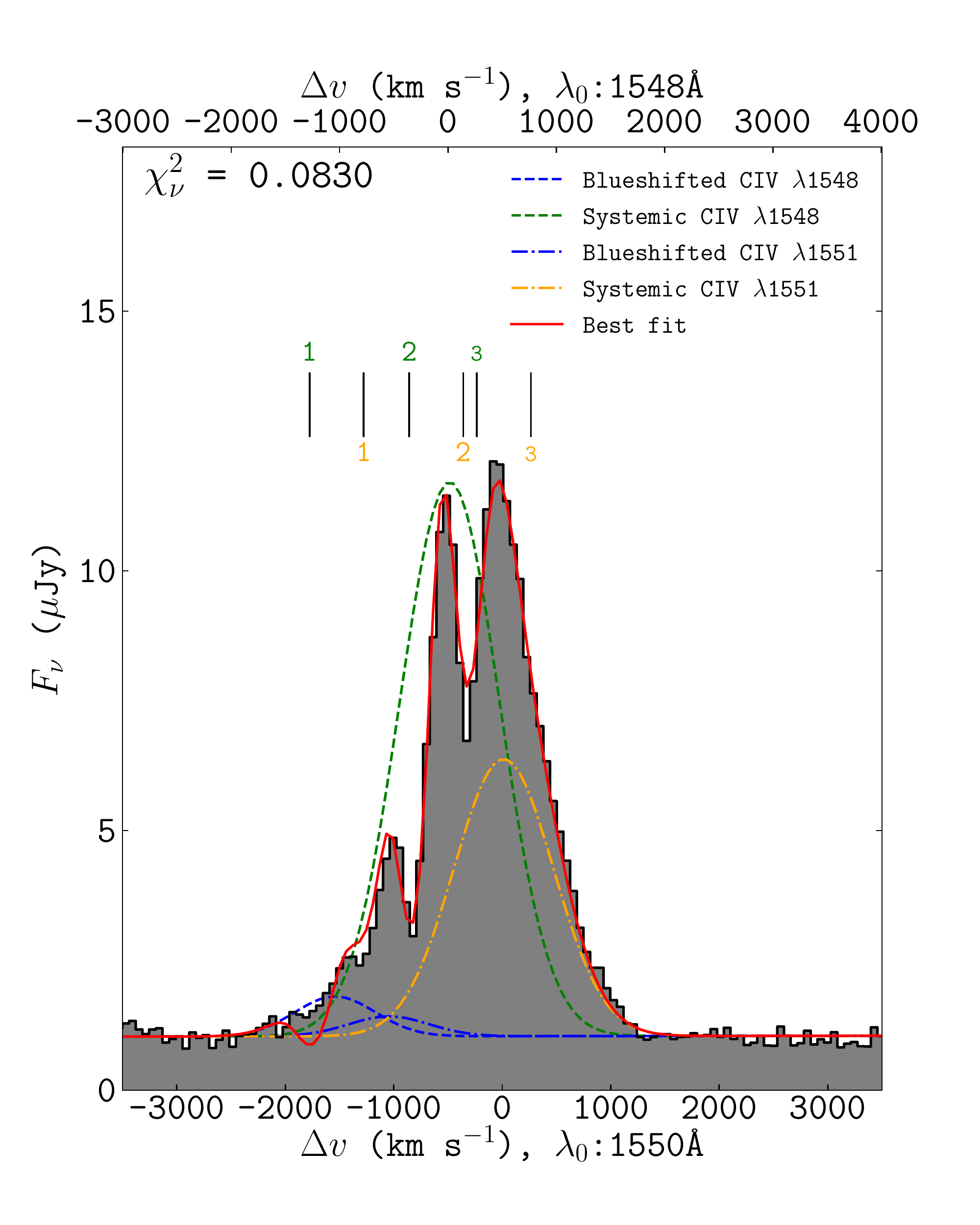}
\caption{Best fit line model to \ion{C}{IV} $\lam\lam1548,1551$ detected with MUSE. The green and orange dashed lines represent the underlying doublet emission at wavelengths of 1548 $\ang$ and 1551 $\ang,$ which are the rest-frame velocities, in the $upper$ and $low$ axes, respectively. Three Voigt profiles model the absorbers for each emission line in the doublet.}
\label{fig:CIV-line}
\end{figure}

\begin{figure} 
\centering
\includegraphics[width=\columnwidth]{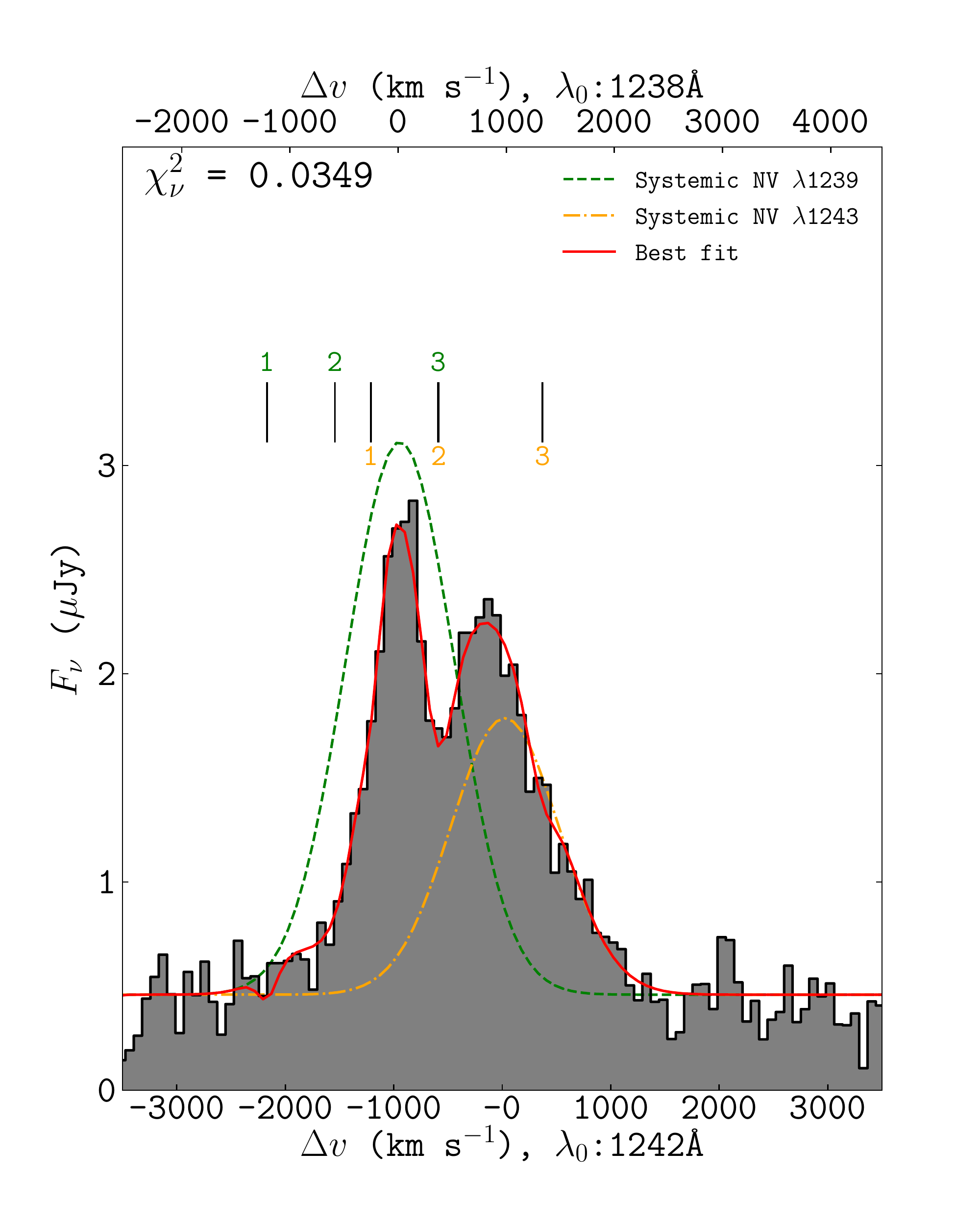}
\caption{Best fit line model to \ion{N}{V} $\lam\lam1238,1242$ detected with MUSE. The green and orange dashed lines represent the underlying doublet emission at the rest-frame wavelengths, 1238 and 1242 $\ang$ which are fixed to the systemic velocity in the $upper$ and $low$ axes. Three Voigt profiles model the absorbers for each emission line in the doublet. }
\label{fig:NV-line}
\end{figure}

The best fit models for \ion{C}{IV} and \ion{N}{V} are shown in Figs~\ref{fig:CIV-line} and \ref{fig:NV-line}. The emission and absorption fit results are shown in Tables \ref{table:absorption-fits} and \ref{table:emission-fits}, respectively. We note that both \ion{C}{IV} and \ion{N}{V} fits feature a slight kink at $\Delta \varv$ $\sim -2000$ km s$^{-1}.$ This may be result of the column density of absorber 1 being over-estimated at this velocity which could imply that the blueshifted \ion{C}{IV} emission is not impeded by absorber 1, as we have assumed. Rather, absorber 1 covers the emission line region behind it the blueshifted emission. Determining which absorbers cover the systemic and/or the blueshifted emission is a task that will require higher spatial and/or spectral resolution. For simplicity, we assume that all three absorbers impede the systemic, blueshifted and continuum emission components.

\begin{figure} 
\centering
\includegraphics[width=\columnwidth]{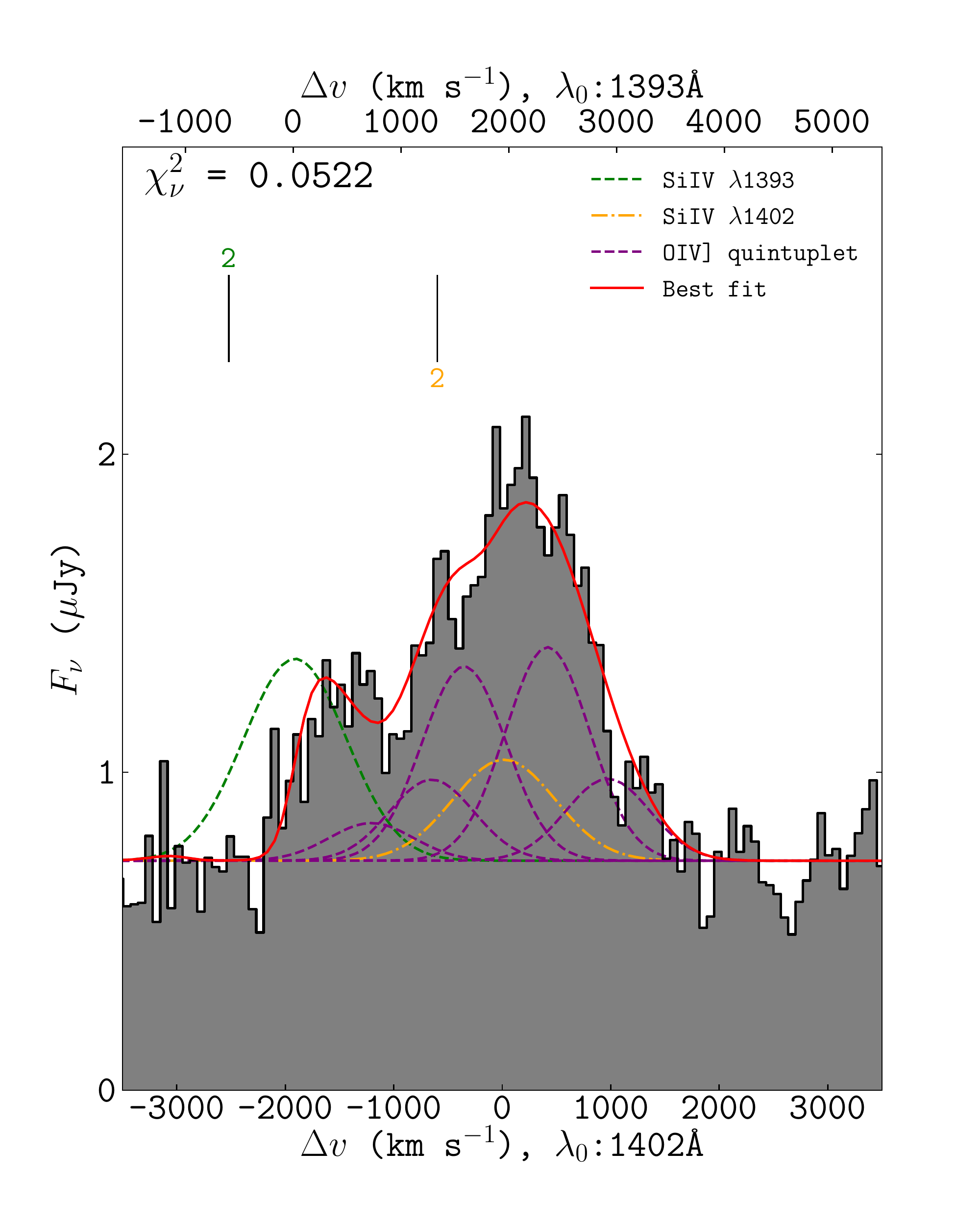}
\caption{Best fit line model to the intercombination \ion{Si}{IV} $\lam\lam1393,1402$ + \ion{O}{IV]} line in MUSE. The green and orange dashed lines represent the underlying doublet emission at the rest-frame wavelengths, 1393 and 1402 $\ang$ which are fixed to the systemic velocity in the $upper$ and $low$ axes. The \ion{O}{IV]} quintuplet line emission is shown in purple.}
\label{fig:SiIV-line}
\end{figure}

\subsection{\ion{Si}{IV}+\ion{O}{IV}]}\label{section:SiIV-fit}

The detected \ion{Si}{IV} $\lam\lam1393,1402$ line doublet overlaps with emission from the \ion{O}{IV]} quintuplet. \ion{Si}{IV} is a resonance line and we model it with an absorption line at the same velocity as \lya absorber 2. We do not include absorbers 1 and 3 because the fit does not change significantly when they are added.  

As before, the Voigt profile is convolved with the LSF of MUSE and the \ion{Si}{IV} doublet emission  is modelled by two Gaussians. Again, we have fixed the continuum to the result of the linear polynomial fit of the local continuum (while the line emission was masked). The \ion{O}{IV]} quintuplet comprises five emission components at the rest-frame wavelengths for \ion{O}{IV]} which are 1397.2 $\ang$, 1399.8 $\ang$, 1401.2 $\ang$, 1404.8 $\ang$, 1407.4 $\ang.$ To fit these, we use five Gaussian profiles with equal line widths. The quintuplet ratios are set by the oscillator strengths of each transition. We show the fit for the intercombination \ion{Si}{IV}+\ion{O}{IV]} lines in Fig.~\ref{fig:SiIV-line}, with the absorption fit results shown in Table \ref{table:absorption-fits} and those for emission shown in Table \ref{table:emission-fits}. 

In an attempt to confirm the consistency of this result, we compare the fluxes of the same \ion{Si}{IV} and \ion{O}{IV]} lines detected at high spectral resolution for the binary symbiotic star, RR Telescopii in shown in Fig. 5 of \citet{keenan2002}. The best fit flux ratios for \ion{O}{IV]} are consistent with those of the high resolution stellar spectrum, which is the perhaps the best observational comparison available for \ion{Si}{IV} and \ion{O}{IV]} intercombination line fluxes. 

\begin{table*}
\caption{Lower and upper bounds placed on initial conditions in the line-fitting routine.
\newline {\it Note:}$\Delta$ is the largest permissible deviation (from the initial guess) imposed on a fit parameter.}
\centering
\begin{tabular}{ l  l }
\hline \hline
Fit parameters & Boundary conditions \\
														& \\
  \hline
														& \\
\bf{Gaussian (for emission):} 			& \\
														& \\
Line centre, $\lam_0$ (\ang) 					& $\Delta \lam_0 = 0.3$ \\
Line flux, $F$ (erg s$^{-1}$ cm$^{-2}$)	& $F \geq$ 0 \\
Line width, $\sigma_\lam$ (\ang) 			& $\sigma_\lam > 0$ \\								
														& \\
 \hline
														& \\
\bf{Voigt (for absorption):} 				& \\
														& \\
\underline{Redshift, $z$:}  				& \\	
 \lya 													&	$\Delta z_1 \leq 5.0\e{-3}$  \\
 														&	$\Delta z_2 \leq 1.0\e{-3}$ \\
 														&	$\Delta z_3 \leq 4.0\e{-3}$ \\
														&	$\Delta z_4 \leq 3.0\e{-3}$ \\
\ion{C}{IV}										& $\Delta z_n \leq 6.0\e{-4}$\\
\ion{Si}{IV} and \ion{N}{V}  				& $\Delta z_n \leq 6.0\e{-3}$\\
																		& \\
\underline{Doppler parameter, $b$ (km s$^{-1}$):} 			& \\																
Ly$\alpha,$ \ion{C}{IV}, \ion{N}{V} and \ion{Si}{IV} 			& 40 $\leq b \leq$  400 \\
 																		& \\
\underline{Column density, $N$ (cm$^{-2}$): }	& \\
\lya 																	& 10$^{12}$ $\leq N \leq$ 10$^{20}$ \\
\ion{C}{IV}, \ion{N}{V} and \ion{Si}{IV} 				& 10$^{13}$ $\leq N \leq$ 10$^{16}$ \\			
																		& \\
  \hline
\end{tabular}
\label{table:absorption-limits}
\end{table*}

\begin{table*}
\caption{The doublet constraints, set by atomic physics, are embedded in the fitting to obtain best fit results for \ion{C}{IV}, \ion{N}{V} and \ion{Si}{IV}+\ion{O}{IV]}. The flux ratios (${F_1}/{F_2}$) are equal the oscillator strengths ratios ($f_1/f_2$). The line centre ratios ($\lam_{0,1}/\lam_{0,2}$) are equal to the rest-frame wavelengths ratios ($\lam_1/\lam_2).$ The redshift, $z,$ Doppler parameter and column density, $N$ are equal between doublet wavelengths. The \ion{O}{IV]} quintuplet constraints are similar to those of the doublets. The fourth quintuplet line, 1404.8 $\ang,$ is the brightest of the \ion{O}{IV]} quintuplets, hence the other four \ion{O}{IV]} lines are fixed to it. }
\centering
\begin{tabular}{ l  l }
\hline \hline
Fit Parameters &  Rules \\
 	& \\
 \hline
  	& \\
 \underline{Gaussian parameters for doublet lines:} 	& \\
 	& \\
Line centre, $\lam_0$ (\ang)							& $\frac{\lam_{0,1}}{\lam_{0,2}} = \frac{\lam_1}{\lam_2}$ \\
Line flux, $F$ (erg s$^{-1}$ cm$^{-2}$)		& $\frac{F_1}{F_2} = \frac{f_1}{f_2}$ \\
Line width, $\sigma_\lam$ (\ang)				& $\sigma_{\lam,1} = \sigma_{\lam,2}$ \\
 	& \\
  \underline{Gaussian parameters for \ion{O}{IV]} quintuplet line ($n = 1, 2, 3, 5$):} & \\
  	& \\
  Line centre,  $\lam_0$ (\ang)						& $\frac{\lam_{0,n}}{\lam_{0,4}} = \frac{\lam_n}{\lam_1}$ \\
  Line flux, $F$ (erg s$^{-1}$ cm$^{-2}$) 	& $\frac{F_n}{F_4} = \frac{f_n}{f_4}$ \\
 Line width, $\sigma_\lam$ (\ang)				& $\sigma_{\lam,4} = \sigma_{\lam,n}$ \\
  	& \\
 \underline{Voigt parameters for multiplet lines with $n$ transitions:} & \\
 	& \\
Redshift, $z$											& $z_1 = z_n$ \\
Doppler parameter, $b$ (km s$^{-1}$)		& $b_1 = b_n$ \\
Column density, $N$ (cm$^{-2}$) 			& $N_1 = N_n$ \\
 	& \\
\hline
\end{tabular}
\label{table:absorption-rules}
\end{table*}

\begin{table*}
\caption{Best fit results to the absorbers in the UVES \lya spectrum from this work and the literature of \citet{jarvis2003} and \citet{wilman2004}.}
\centering
\begin{tabular}{c D{,}{\, \,\pm\, \,}{-3} D{,}{\, \,\pm\, \,}{-3} D{,}{\, \,\pm\, \,}{-3}   }
\hline\hline       		
Absorber 		& \mc{Absorber redshift}	  		& \mc{Column density}   						& \mc{Doppler parameter}   \\
	\#					& z										& \mc{$N_{\ion{H}{I}}$ (cm$^{-2}$)}  	& \mc{$b$ (km s$^{-1}$)}      \\ 
												& & & \\
						\hline
						& & & \\
						UVES  & & & \\
						(this work) & & & \\
		1				& 2.9063,0.0001		& (1.363,0.217) \e{14} 	& 107,15\\
		2				& 2.9185,0.0001   	& (1.262,0.148) \e{19} 	& 58,1 \\
		3				& 2.9262,0.0001   	& (5.166,0.824) \e{13}  & 133,15\\
		4				& 2.9324,0.0001   	& (2.232,0.310) \e{13}  & 25,4 \\
						& & & \\
						UVES  & & & \\
						(literature$^a$) & & & \\
		1				& 	2.9066,0.0062 		& (1.047,0.314) \e{14} 	& 88,45 \\
		2				& 	2.9185,0.0001 		& (1.202,0.072) \e{19} 	& 58,3 \\
		3  				& 	2.9261,0.0005 		& (3.548,0.568) \e{13} 	& 109,35 \\
		4				& 	2.9324,0.0001 		& (2.239,0.672) \e{13} 	& 23,17 \\
						& & & \\
\hline
\end{tabular}
\label{table:absorption-fits-uves}
\end{table*}

\begin{table*}[ht]
\caption[]{Best fit results to the non-absorbed emission in the UVES spectrum. The flux units are arbitrary (arb.). Blueshifted lines are labelled by the abbreviation ``bl.''.}    
\centering                          
\begin{tabular}{ l c D{,}{\, \,\pm\, \,}{-3} D{,}{\, \,\pm\, \,}{-3} D{,}{\, \,\pm\, \,}{-3} }
\hline\hline UVES  \\
\hline        
Line 	&  
\mc{Line centre (rest)} & 
\mc{Line centre (obs.)} 
&\mc{Line flux}  
& \mc{Line width}   \\   
 	&  
\mc{$\lam_0$ (\ang)} & 
\mc{$\lam$ (\ang)} & \mc{$F$ (arb. units)}
&\mc{FWHM (km s$^{-1}$)}  \\
&  &  \mc{} & \mc{} &\mc{}  \\ \hline     
&  &  \mc{} & \mc{} &\mc{}  \\
  \lya							& 1215.67  	& 4769.07,2.76 	& 0.56,0.14 	& 1427.67,82.47 \\    
  \lya (bl.)  	& "  				& 4751.99,16.73 	& 0.01,0.01	& 1525.40,779.05 \\            
	    							&  &  \mc{} & \mc{} &\mc{} \\   
\hline                                   
\end{tabular} 
\label{table:emission-fits-uves}  
\end{table*}

\begin{table*}
\caption{Best fit results to the absorbers in the MUSE spectrum. The uncertainties reported are 1$\sigma$ errorbars. The parameters that are prefixed by $\sim$ are those fit parameters that were fit with very large uncertainties in the least-squares fitting routine. Column (1) indicates the absorber (abs.) number. The column density fit parameters with large uncertainties have been quoted as upper limits. Column (4) is the central wavelength (wav.) of the absorber. Column (8) is the rest-frame equivalent width (E.W.) (for \ion{Si}{II} lines only). 
\newline {\it Note:} $^{a}$ \ion{Si}{II} $\lam1260$ and $^{b}$ \ion{Si}{II} $\lam1526$}
\centering                          
\begin{tabular}{l l D{,}{\, \,\pm\, \,}{-5} D{,}{\, \,\pm\, \,}{-3} D{,}{\, \,\pm\, \,}{-5} D{,}{\, \,\times\, \,}{-2} D{,}{\, \,\pm\, \,}{-5} D{,}{\geq \, \,}{3}}
\hline\hline    
MUSE \\
\hline   	
Abs. 	& 
Ion 		& 
\mc{Redshift}  	& 
\mc{Absorber wav.} &
\mc{Velocity} 	& 
\mc{Column density}   	& 
\mc{Doppler} & 
\mc{E.W.}  \\
\#	& 
 & 
\mc{$z$} 		& 
\mc{$\lam$ (\ang)} &
\mc{$\Delta \varv$ (km s$^{-1}$)} 	& 
\mc{$N$ (cm$^{-2}$)}  								& 
\mc{$b$ (km s$^{-1}$)}  & 
\mc{W$_{\lam,0}$}    \\  
 	&  & 	&  \mc{} & \mc{} &\mc{} \\   \hline
 	&  & 	&  \mc{} & \mc{} &\mc{} \\
1	& \lya 				& 2.9063,0.0003 	& 4748.74,0.51			& -1315,672			& (1.64 \pm 0.52),10^{14} 		& 149,44  \\
	& \ion{N}{V} 		& 2.9076,0.0008	& 4840.85,1.39 		& -1212,1686		& \leq1.04,10^{14}					& \sim101 \\
  	& \ion{C}{IV}  	& 2.9068,0.0007	& 6048.40 				& -1278,1915 		& \leq3.78,10^{14} 					& 197,93 \\
	&  					& 							&								&  \mc{} 				& \mc{} 									& \mc{} \\   
2 	& \lya 				& 2.9184,0.0002 	&	4763.54,0.28 		& -385,109 			& (1.63 \pm 0.46),10^{19}  		& 45,31 	 \\  
 	& \ion{N}{V}		& 2.9158,0.0008 	& 4851.01,1.26			& -585,740			& (9.40 \pm 4.10),10^{14}			& 365,78  \\
 	& \ion{Si}{IV} 	& \sim2.9157 		& 5457.49 				& \sim-598 			& \leq2.40,10^{15}					& \sim400  \\
 	& \ion{C}{IV} 	& 2.9188,0.0001 	& 6067.05 				& -357,52				& (4.52 \pm 1.29),10^{14} 		& 162,22 \\
 	& \ion{Si}{II}$^a$  	& 2.9212,0.0001  	& 4942.28,1.67		&  -175,293			& \geq1.83\e{14} 						& 	& ,0.79 \\   
	& \ion{Si}{II}$^b$ 	& 2.9183,0.0006  	& 5982.04,1.16		&  -397,461			& \geq5.35\e{14}  					&  & ,0.41 \\  
	&  					& 							&								&  \mc{} 				& \mc{} 									& \mc{} \\   
3 & \lya 				& 2.9273,0.0009 	& 4774.32,1.41			& 293,412 			& (2.06 \pm 1.54),10^{13} 		& \sim104 \\
   	& \ion{N}{V} 		& 2.9283,0.0006 	& 4866.50 				& 370,363 			& \leq1.61,10^{14} 					& 157,71\\
   	& \ion{C}{IV}  	& 2.9267,0.0006 	& 6079.31 				& 248,297				& \leq6.03,10^{13}  					& 212,81 \\ 
	&  					& 							&								&  \mc{} 				& \mc{} 									& \mc{} \\     
4 & \lya 				& 2.9321,0.0004	& 4780.14,0.67 		& 658,442 			& (2.07 \pm 1.28),10^{13} 		& \sim50 \\
	&  					& 							&								&  \mc{} 				& \mc{} 									& \mc{} \\   
\hline
\end{tabular}
\label{table:absorption-fits}  
\end{table*}

\begin{table*} 
\caption[]{Best fit results to non-absorbed emission in the MUSE spectrum. The uncertainties shown are 1$\sigma$ errorbars. Values prefixed by $\sim$ are results that were fit with very large systemic uncertainties (as in Table \ref{table:absorption-fits}). Blueshifted lines are labelled by the abbreviation ``bl.''.}    
\centering                          
\begin{tabular}{l c D{,}{\, \,\pm\, \,}{-3} D{,}{\, \,\pm\, \,}{-3} D{,}{\, \,\pm\, \,}{-3} }
\hline\hline                                                                         MUSE  \\
\hline            
Ion 	&  
\mc{Line centre (rest)} & 
\mc{Line centre (obs.)} 
&\mc{Line flux}  
& \mc{Line width }   \\    
 	&  
\mc{$\lam_0$ (\ang)} & 
\mc{$\lam$ (\ang)} 
&\mc{$F$ (10$^{-17}$ erg s$^{-1}$ cm$^{-2}$)}  
& \mc{FWHM (km s$^{-1}$)}   \\   
	    	&  &  \mc{} & \mc{} &\mc{} \\   
\hline 
	    	&  &  \mc{} & \mc{} &\mc{} \\   
   \lya										& 1215.67  	& 4769.50,2.65  			& 158.5,63.8 	& 1511,108  \\    
   \lya (bl.)   					& "  				& 4752.23,57.14 			& \sim11.9 		& \sim1302 \\  
   \ion{N}{V}								& 1238.82 	& 4860.80,1.37 			& 6.8,0.9			& 1180,87 \\
   												& 1242.80 	& 4876.49,1.37 			& 3.4,0.5			& 1175,87\\ 
   \ion{Si}{IV}								& 1393.76 	& \sim5468.70				& \sim1.4			& \sim1118  \\
   												& 1402.77 	& \sim5504.05 				& \sim0.7 			& \sim1111 \\
   \ion{O}{IV]}							& 1397.20 	& \sim5481.60 				& \sim0.2 			& \sim920 \\
   												& 1399.80    & \sim5491.80				& \sim0.5			& \sim918 \\
   												& 1401.20		& \sim5497.29				& \sim1.1			& \sim917 \\
   												& 1404.80		& \sim5511.42  			& \sim1.2			& \sim915 \\
   												& 1407.40		& \sim5521.62 				& \sim0.5			& \sim913 \\
   \ion{C}{IV} 							& 1548.20  	& 6074.59,1.78 			& 20.4,3.1			& 1090,99   \\ 
   \ion{C}{IV} (bl.) 		& " 				& 6053.16,20.43			& \leq1.2				& \sim877 \\
   \ion{C}{IV}								& 1550.77		& 6084.68,1.78 			& 10.2,1.5 		& 1088,99  \\
   \ion{C}{IV} (bl.) 		& " 				& 6063.21,20.39			&	\leq0.6				& \sim756 \\ 									
   \ion{He}{II}       						& 1640.40		& 6436.09,0.30 			& 16.4,0.5			& 978,24 	\\ 
   \ion{He}{II} (bl.)  		& " 				& 6413.49,3.05  			& 1.4,0.5 			& 970,225  \\           
   \ion{C}{III]}								& 1906.7		& \sim7481.21				& \sim7.4			& \sim977 \\
   \ion{C}{III]} (bl.)			& "				& \sim7454.92				& \leq0.5				& \sim947 \\
   \ion{C}{III]}								& 1908.7		& \sim7489.06				& \sim3.7			& \sim976 \\
   \ion{C}{III]} (bl.)			& "   				& \sim7462.74			   	& \leq0.2				& \sim946 \\
   \ion{C}{II]} 								& 2326.9		& \sim9129.26				& 4.0				& \sim1300 \\
   \ion{C}{II]} (bl.)			& " 				& \sim9096.15				& 1.2		    		& \sim1152 \\		
    	&  &  \mc{} & \mc{} &\mc{} \\   
\hline                                   
\end{tabular} 
\label{table:emission-fits}  
\end{table*}

\subsection{\ion{Si}{II}}\label{section:SiII-fit}

\begin{figure}
\centering 
\includegraphics[width=0.45\textwidth]{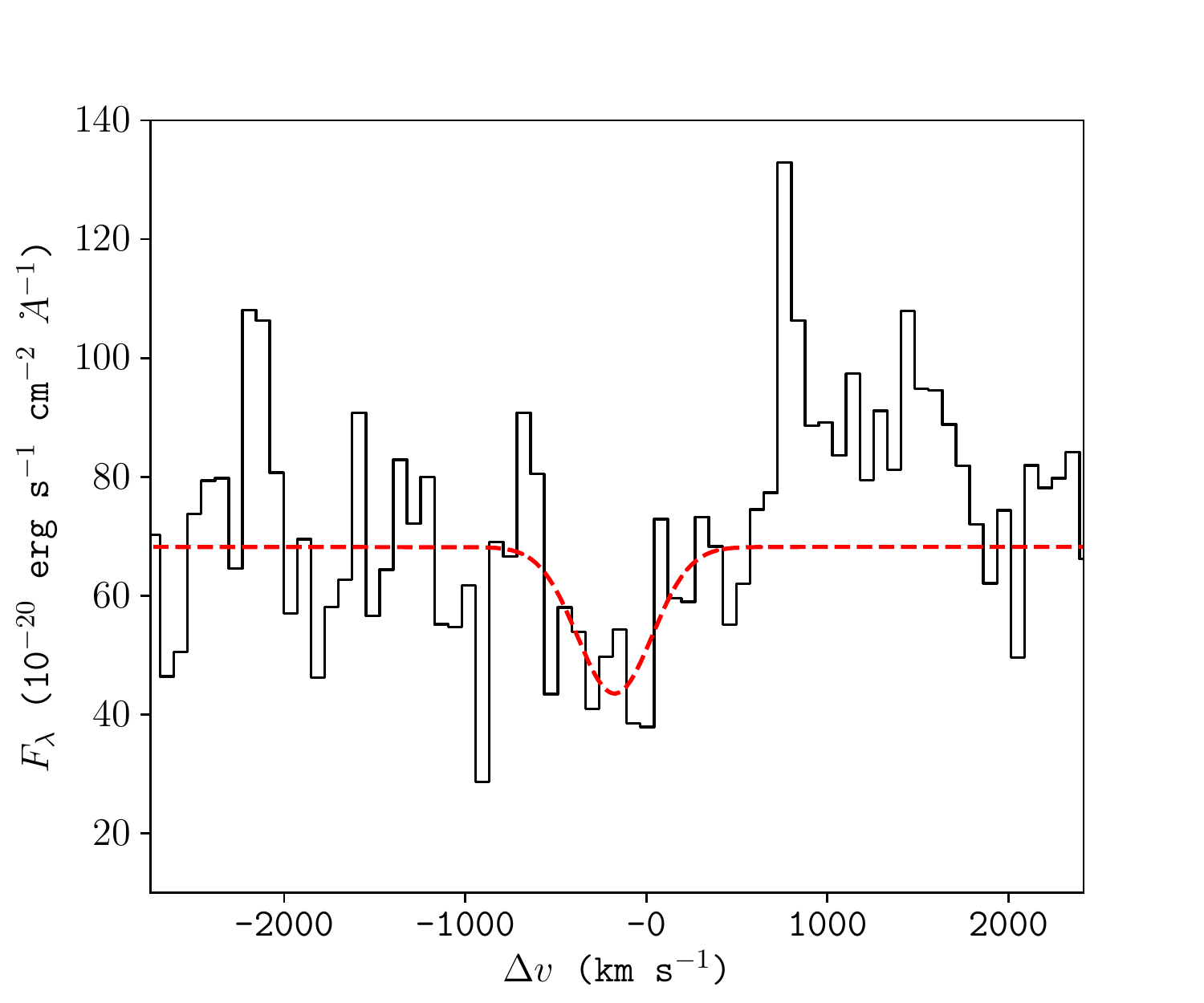}
\caption{\ion{Si}{II} $\lam$1260 absorption line fit with a single Gaussian component with the best fit line shown in red.  }
\label{fig:SiII_1260-fit}
\end{figure}

\begin{figure}
\centering 
\includegraphics[width=0.45\textwidth]{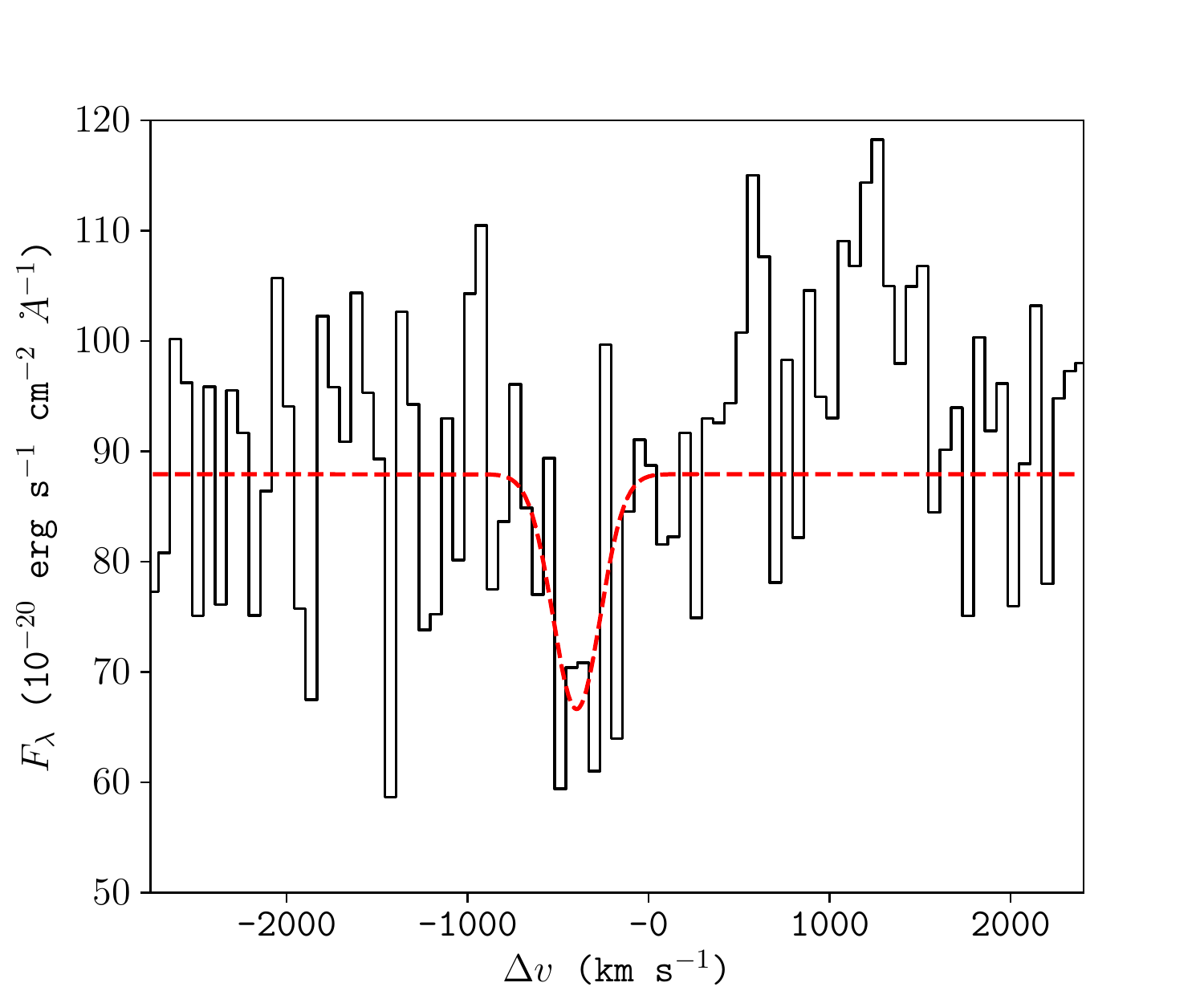}
\caption{\ion{Si}{II} $\lam$1527 absorption line fit with a single Gaussian component with the best fit line shown in red. }
\label{fig:SiII_1526-fit}
\end{figure}

We have detected \ion{Si}{II} $\lam$1260 and \ion{Si}{II} $\lam$1527 absorption in the rest-frame UV spectrum which are fit with Gaussians to account for the absorbed components (see Figs \ref{fig:SiII_1260-fit} and \ref{fig:SiII_1526-fit}). Using this fit, we estimate their velocity shifts and column densities using the approximation from \citet{humphrey2008b}, 
\begin{equation}
N \geq \frac { W_\lam m_e c^2 } { \pi e^2 f \lam_0^2 },
\end{equation}
where $N$ is the column density, W$_\lam$ is the observed equivalent width, $m_e$ is the electron mass, $c$ is the light-speed, $e$ is the electron mass, $f$ the oscillator strength and $\lam_0$ the rest wavelength of the line. The column densities are lower limits because it is not possible to determine whether the lines are in the linear or logarithmic (flat) part of the curve of growth.

\begin{figure*} 
\centering
 \includegraphics[width=0.6\textwidth]{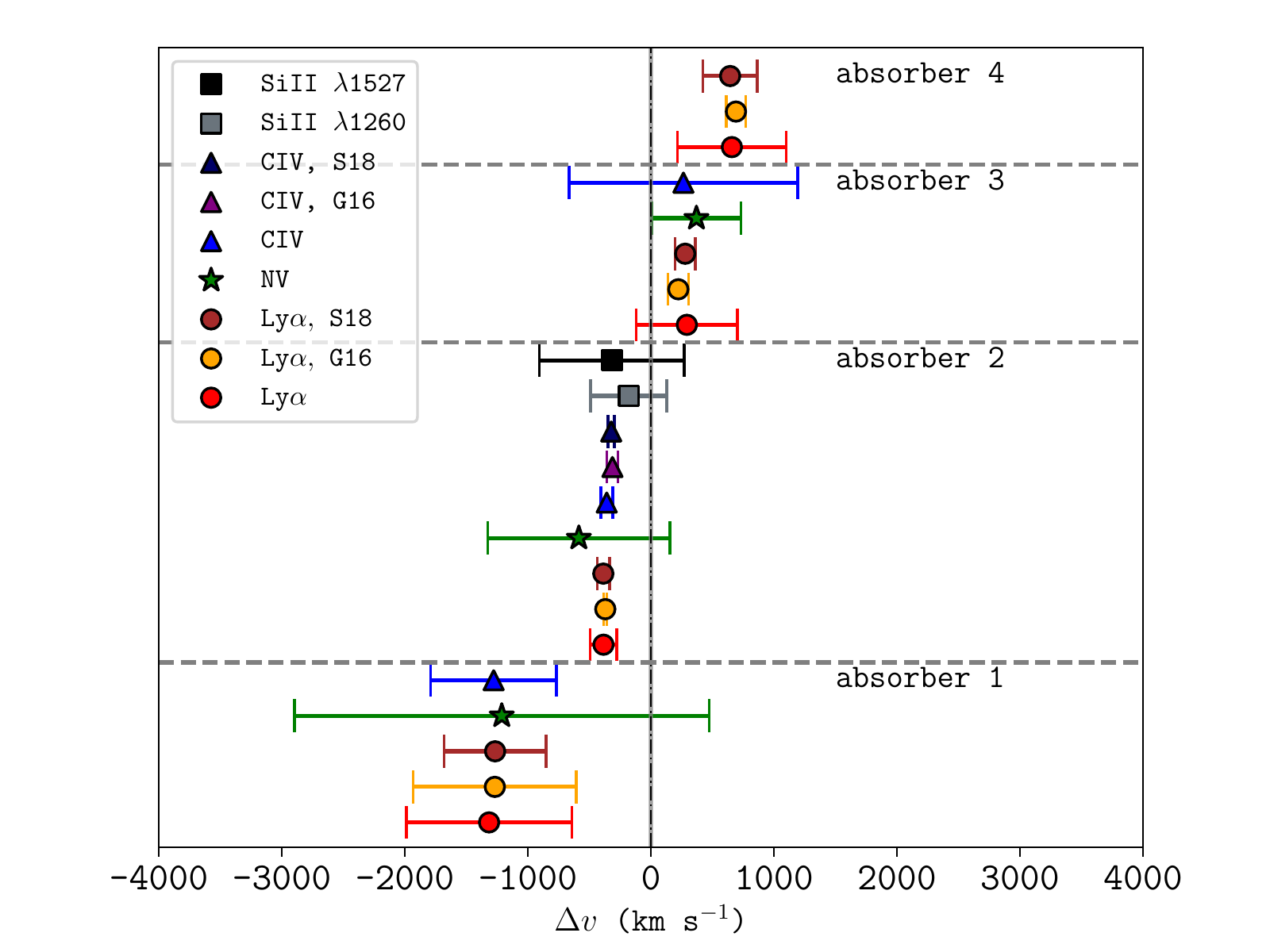}
 \caption{Relative velocities of Ly$\alpha,$ \ion{N}{V} and \ion{C}{IV} and \ion{Si}{II} absorbers from this work as well as those from G16 and S18. The vertical dashed-dotted (black) line indicates the systemic velocity and its error (shaded grey). The horizontal dotted lines (grey) distinguish between the different absorbers.}
 \label{fig:abs-vel}
\end{figure*}

The column densities of the \ion{Si}{II} lines place them in the category of weak absorbers. Their velocity shifts are in agreement with that of \lya absorber 2 meaning that these absorptions also occur within roughly the same gas volume as those of \ion{H}{I}, \ion{C}{IV}, \ion{N}{V}, \ion{Si}{IV} absorbers (as Fig. \ref{fig:abs-vel} shows). One implication of this finding is that it is unlikely for the strong absorber to be matter bounded\footnote{A matter bounded cloud is insufficiently optically thick to absorb all of the incident UV photons.}. For absorber 2 to be matter-bounded, low ionisation species such as \ion{Si}{II} would exist in only trace amounts compared to higher ionisation lines. A clear detection of both these ions at the same velocity as the strong absorber proves that the absorber is more probably ionisation bounded and therefore unity covering factor. This also implies that ionising photons will not be able to escape from the CGM of this source and that ionising radiation emerging emerging from the halo will not contribute significantly to the metagalactic background or ionisation of gas in intergalactic medium (IGM).

\subsection{\ion{C}{III]} and \ion{C}{II]}}
\begin{figure} 
\centering
 \includegraphics[width=\columnwidth]{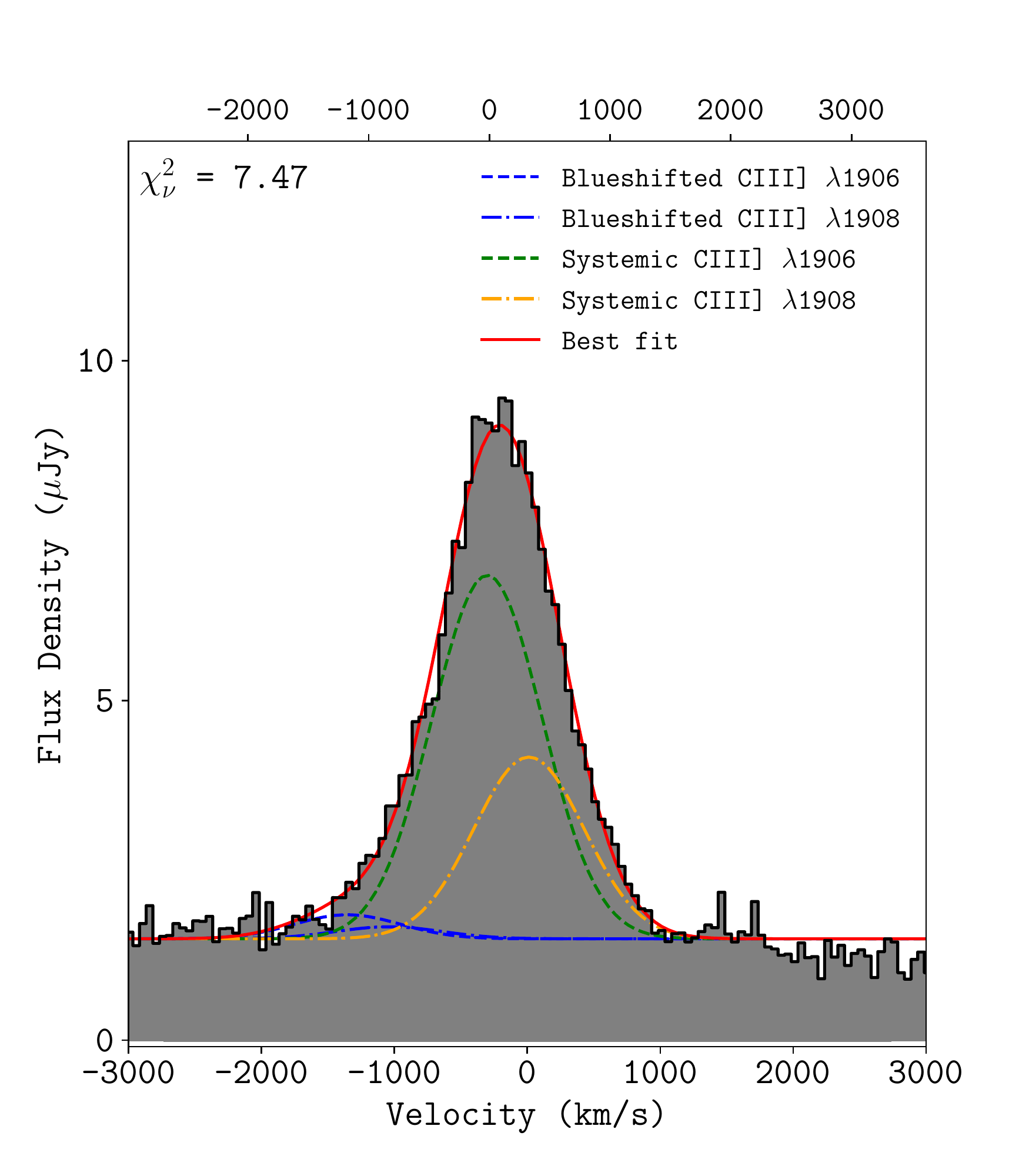}
 \caption{Best fit line model to \ion{C}{III]} $\lam\lam1906,1908$ doublet that includes a blueshifted emission component.}
 \label{fig:CIII-fit}
\end{figure}

\begin{figure} 
\centering
 \includegraphics[width=\columnwidth]{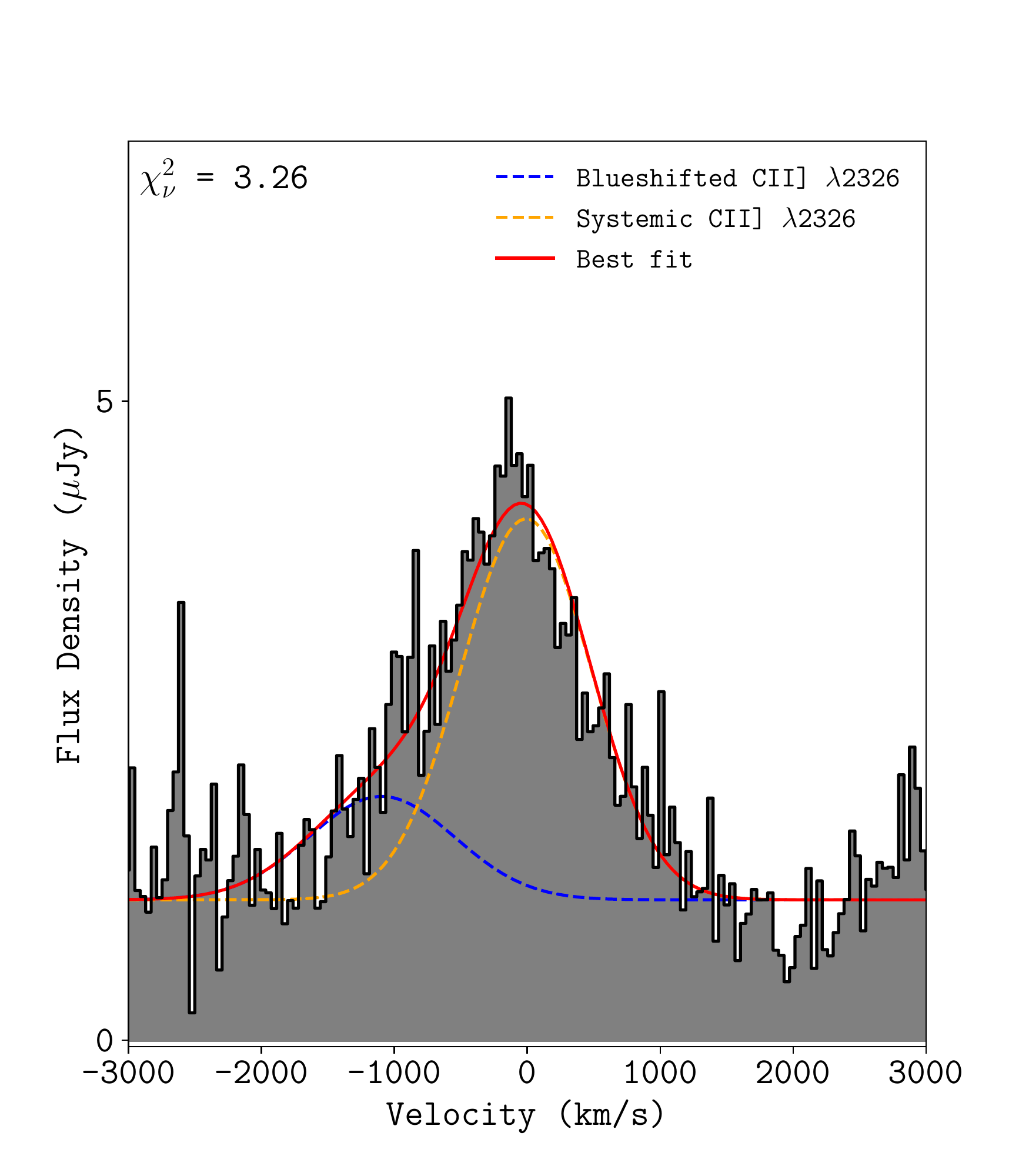}
 \caption{Best fit line model to \ion{C}{II]} $\lam2326$ singlet that includes a blueshifted emission component.}
 \label{fig:CII-fit}
\end{figure}

The gas tracers \ion{C}{III]} $\lam\lam1906,1908$ and \ion{C}{II]} $\lam2326$ are non-resonant lines but useful tracers when searching for evidence of shock ionisation which we discuss in more detail in section \ref{section:photoionisation-modelling}. Hence, we have included them in the line-fitting routine. To the \ion{C}{III]} doublet, we fit four Gaussian components in total to account for emission from the doublet at both systemic and blueshifted velocities as we have done for Ly$\alpha,$ \ion{He}{II} and \ion{C}{IV}. \ion{C}{III]} has a sufficiently high surface brightness for us to fit it in the same way. Its best fit result is shown in Fig. \ref{fig:CIII-fit}. \ion{C}{II]} is a singlet which showed evidence of very blueshifted emission relative to systemic and thus we have also added an additional Gaussian component when fitting it (see Fig. \ref{fig:CII-fit}). The best fit emission parameters for both of these lines are shown in Table \ref{table:emission-fits}.

\section{Morphology of the absorbers}\label{section:morphology-absorbers}
The morphology of the absorbers within the circumgalactic medium of Yggdrasil are a focus of interest because much uncertainty about their origin and probable fate still remains. Although the absorbers have been studied extensively using data from long-slit \citep[i.e.][]{rottgering1995,vanojik1997} and echelle spectroscopy \citep[i.e.][]{jarvis2003,wilman2004} which were limited in their ability to provide a spatially resolved view of the gas in emission and absorption around Yggdrasil. For instance, they were not capable of showing the spatial variation in the \lya profile which indicates variation in the kinematics of the most extended absorber (absorber 2). 

The MUSE data we have used to perform resonance line-fitting, above, provides us with the capability of estimating the full extent and shape of \lya absorber 2 which has a covering factor of C $\simeq$ 1.0 as G16 and S18 have both shown. Furthermore, we can deduce its neutral and also ionised gas mass thus estimating its total hydrogen gas mass. 

\subsection{Size, shape, mass and ionisation of the strongest \lya absorber}\label{section:morphology}

In agreement with previous work on the \lya line in Yggdrasil, absorber 2, located at a velocity shift of $\Delta \varv \sim-400$ km s$^{-1},$ reaches zero flux at its line centre. This implies that it is saturated with a unity covering factor i.e. C $\simeq$ 1.0 and \lya column density of $\simeq$ $10^{19}$ cm$^{-2}$ (see Table \ref{table:absorption-fits}). In the UVES spectrum, the \lya absorbers occupy disparate velocities. This implies that the structure of the \ion{H}{I} gas is likely to be shell-like as \citet{binette2000} and \citet{jarvis2003} have suggested. Guided by these previous findings and using more recent IFU data, we can estimate the size, shape and mass of absorber 2 which is the strongest \lya absorber. 

In terms of location, we rely on the result given in \citet{binette2000} which showed that the absorbing and emitting gas are not co-spatial. Rather absorber 2 is further out from the halo and screens the radiation from the extended emission line region. We base the rest of our description of the halo gas on this predication. The spatial extent of the absorbing \ion{H}{I} gas medium in \citet{rottgering1995} is found to be $r \gtrsim 13$ kpc based on the extent of \lya emission measured and the assumption of a unity covering factor for the associated absorption. In G16, visual inspection of the IFU data leads to a value of $r \gtrsim 60$ kpc in radial extent. In S18, a velocity gradient across the halo is measured and used to estimate the absorber size which they find to be $r \gtrsim 38$ kpc in radius. In this work, we use the radial size of \ion{H}{I} gas shell from G16 who determined the size by pinpointing the furthest spaxels from the nucleus of the gas halo, at several position angles, where the \lya absorber 2 is still observed. 

In our data, we find that the absorber extends to projected distances of between $r = 50$ kpc and $r = 60$ kpc from the high surface brightness region (HSBR) and is non-isotropic, covering the extended emission line region (EELR) over an area of 50 $\times$ 60 kpc$^{2}.$ A low surface brightness halo with quiescent kinematics (FWHM = 400 - 600 km s$^{-1}$) extending out to $r = 67$ kpc has been detected in this source \citep[e.g.,][]{villar-martin2003}. Ly$\alpha,$ \ion{He}{II}, \ion{N}{V} and \ion{C}{IV} emission are detected in the giant halo. In fact, \ion{N}{V} appears to be strengthened more so than in the quiescent haloes of other HzRGs in the \citet{villar-martin2003}. Given the size of the absorber, it is likely that it covers the emission from the giant halo. 

Assuming spherical symmetry and density homogeneity of the absorbing gas shell, the \ion{H}{I} mass is estimated by $M_\ion{H}{I} = 4 \pi r^2 {\rm m}_\ion{H}{I} N_\ion{H}{I}$ \citep{debreuck2003,humphrey2008b}. Taking into account the estimated size of $r \gtrsim 60$ kpc and an \ion{H}{I} column density of $10^{19.2}$ cm$^{-2}$ (from Table \ref{table:absorption-fits}), the \ion{H}{I} mass of absorber 2 is $M_\ion{H}{I}/M_\odot = 5.7\e{9} (r / 60 {\rm kpc})^2$ $(N_\ion{H}{I} / 10^{19.2} {\rm cm}^{-2}).$ Our results are within, at most, two orders of magnitude agreement with those in the literature. In \citet{rottgering1995}, the mass of absorber 2 is estimated as, $ M_\ion{H}{I}/M_\odot \gtrsim 2.0\e{7} (N_\ion{H}{I} / 10^{19} {\rm cm}^{-2})(r / 13 {\rm kpc}).$ In G16, this value is $M_\ion{H}{I}/M_\odot \gtrsim 3.8\e{9}(r / 60 {\rm kpc})^2$ $(N_\ion{H}{I} / 10^{19} {\rm cm}^{-2}) $ and the estimate given in S18 of $M_\ion{H}{I}/M_\odot \gtrsim 10^{8.3}.$

An approximation of the hydrogen ionisation fraction X$_\ion{H}{II} = \ion{H}{II}/(\ion{H}{I}+\ion{H}{II})$ is clearly required to estimate the gas mass of the absorbing structure from our observational measurement of $N_\ion{H}{I}.$ In principle, the value of X$_\ion{H}{II}$ can vary from zero in the case of purely neutral gas, to $\sim$1.0 in the case of matter-bounded, photoionised gas \citep[e.g.,][]{binette1996,wilson1997}. In the absence of a method to directly estimate N$_\ion{H}{II},$ we instead turn to the carbon ionisation fraction, X$_{\rm C}$, which we define here as the ratio of all ionised species of carbon to all species of atomic carbon (i.e., ionised or neutral). In the case of Yggdrasil, we combine the measurement of $N_\ion{C}{IV}$ with the upper limit $N_\ion{C}{I}$ $\le$ 1.5 $\times$10$^{14}$ cm$^{-2}$ from S18, to obtain X$_{\rm C}$ $\ge$ $N_\ion{C}{IV} / (N_\ion{C}{I}+N_\ion{C}{IV}) = 0.8.$ 

Note that this is a lower limit because we have no useful constraints on any of the other ionised species of carbon. The fact that the ionisation energy of \ion{C}{I} (11.3 eV) is similar to that of \ion{H}{I} (13.6 eV) means that under photoionisation we can assume that the ionisation fraction of hydrogen and carbon are similar i.e. X$_\ion{H}{II}$ $\sim$ X$_{\rm C},$ and thus we obtain X$_\ion{H}{II}$ $\ga$ 0.8. This falls within the framework of the absorber being ionisation rather than matter bounded suggested by the \ion{Si}{II} detections in section \ref{section:SiII-fit}.  

The neutral fraction is what is remaining i.e. X$_\ion{H}{I} \lesssim 0.2.$ Assuming that the absorber is a two-phase medium as in \citet{binette2000}, the ionised fraction implies that $M_\ion{H}{II}/M_\ion{H}{I} \ga 4.$ Using this, we can estimate the total hydrogen mass (excluding the molecular gas contribution) of the absorber such that it is, $M_{\rm T}/M_\odot \gtrsim M_\ion{H}{I} + M_\ion{H}{II} = 5 M_\ion{H}{I}$ hence $M_{\rm T}/M_\odot \gtrsim 2.9\e{10}.$ Absorber 2 is approximately an order of magnitude lower than the stellar-mass of the host galaxy which is $M_*/M_\odot = 1.2\e{11}$ \citep{seymour2007}. 

\subsection{Arrangement of the absorbers along the line-of-sight}\label{section:arrangement-absorbers}

Fig. \ref{fig:abs-vel} shows that \lya absorption occurs in the same gas volume as absorbers 1, 2 and 3 in \ion{C}{IV}, \ion{N}{V} and \ion{Si}{IV}. We can determine, from the velocities of these absorbers, a) their geometric arrangement in the halo and b) their kinematics. Are these outflowing \citep[e.g.,][]{zirm2005,nesvadba2006} or infalling gas \citep[e.g.,][]{BarkanaLoeb2003,humphrey2008b}?

In agreement with general observations of HzRGs, \lya absorbers 1 and 2 are blueshifted relative to the systemic velocity \citep{wilman2004}. According to velocity gradient measures across the diameter of the \lya halo in S18, \lya absorber 2 is outflowing. Assuming that absorbers 1, 2 and 3 are outflowing, we can estimate roughly where they are located in relation to one another along the line-of-sight. At $\Delta \varv \sim-1000$ km s$^{-1},$ absorber 1 has the highest blueshift and is thus more likely to be found at a closer proximity to the nucleus than absorber 2 which is located at $\Delta \varv \sim-400$ km s$^{-1}.$ In other words, absorber 1 has the fastest outflow velocity and is therefore closer the centre of the halo. Absorber 3 is redshifted relative to the systemic velocity and it may be infalling. To measure the distances, we require simulated abundances produced by photoionisation models (shown in section \ref{section:photoionisation-modelling}). 

\section{Blueshifted \ion{He}{II}, \lya and \ion{C}{IV} emission}\label{method:blueshifted-emission}

The blueshifted components in the bright lines \ion{He}{II}, Ly$\alpha$ and \ion{C}{IV} are a clear indication of perturbed gas either in the foreground of systemic emission in the halo or outflowing along the line-of-sight. 

\begin{figure*}
\centering
\includegraphics[width=\textwidth]{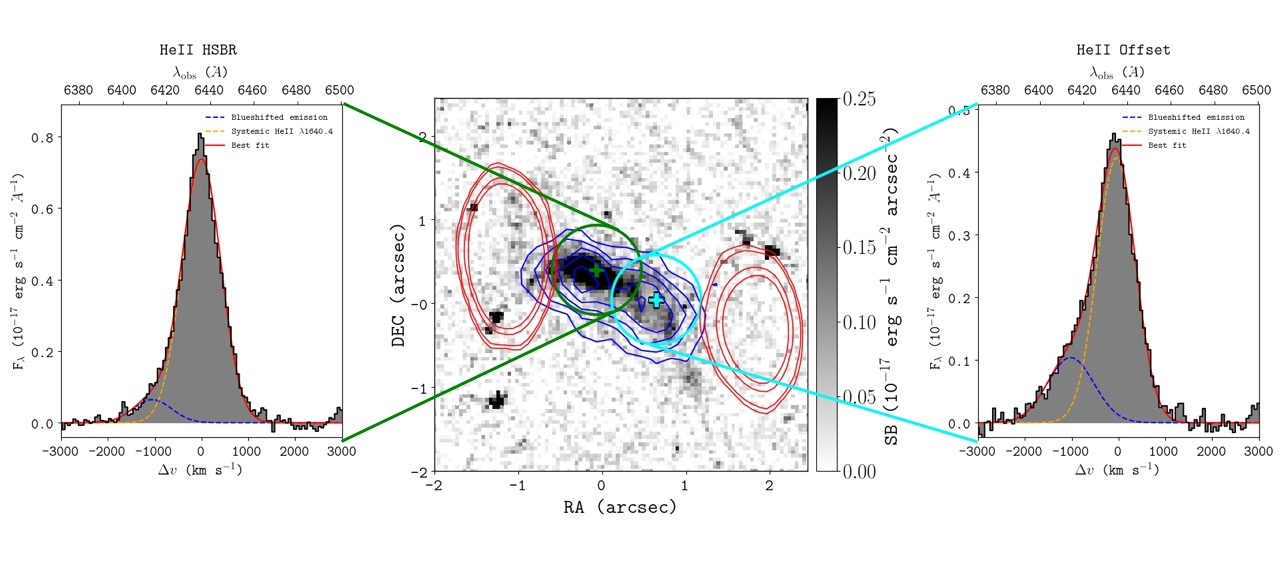}
\caption{A broad-band image of the rest-frame UV continuum of MRC 0942-242 from the {\it HST} WFPC2 702W ($5800 - 8600$ $\ang$) filter is shown. MUSE contours (blue) are overlaid and represent narrow-band emission summed over $6400 - 6425$ $\ang.$ The contours are shown at the surface brightness (SB) levels: (1.0, 1.6, 2.2, 2.8, 3.4, 4.0) $\times 10^{-17}$ erg s$^{-1}$ cm$^{-2}$ $ {\rm arcsec}^{-2}.$ VLA 4.7 GHz radio surface brightness is shown by the red contour levels. The inset 1D spectrums show the high surface brightness region (HSBR) ($left$) and offset ($right$) \ion{He}{II} emission. The line profiles are extracted from apertures of radius, R = 0.6\arcsec, shown in green for the HSBR and cyan for the offset region with the aperture centroids for each shown in the matching colours. The blueshifted and systemic \ion{He}{II} emission are shown in blue and orange, respectively, with the sum of both components shown in red. }
\label{fig:hst-img-muse-contours}
\end{figure*}

To disentangle to the blueshifted and systemic emission in \ion{He}{II}, we use narrow-band imaging. 
A MUSE narrow band image summed over a wavelength range of $6400 - 6425$ $\ang$ is shown as contours in Fig. \ref{fig:hst-img-muse-contours}. Note that the contours are continuum-free. Within this chosen wavelength range, we obtain detections from both the blueshifted and systemic \ion{He}{II} emitters. The image clearly shows two spatially unresolved components: one at the high surface brightness peak and the other spatially offset. 

In addition to this spatial separation, the \ion{He}{II} spectrums extracted from each of these two regions shows that emission at the high-surface brightness region (HSBR, shown in Fig. \ref{fig:0943-continuum}) is dominated by emission from the systemic velocity with trace amounts being blueshifted. The blueshifted component, however, undergoes a flux enhancement by an approximate factor of 2 (see Table \ref{table:HeII-nucleus-offset}) at a projected distance of $r = 1.00 \pm 0.04 \arcsec = 8.0 \pm 0.3$ kpc south-west (PA $\simeq 225^\circ$) from the high surface brightness region (HSBR). 

For further understanding, we show the MUSE \ion{He}{II} contours over a UV/optical {\it Hubble Space Telescope} (HST) Wide Field Planetary Camera 2 (WFPC2) 702W broad-band image. The UV emission detected over $5800 - 8600$ $\ang$ appears to have a bent morphology that extends in the same direction as the region where \ion{He}{II} is enhanced. Also, both the UV broad-band detection and the blueshifted \ion{He}{II} are in alignment with the radio axis detected by 4.7 GHz Very Large Array (VLA) observations \citep{carilli1997,pentericci1999}. 

\begin{figure*}
\centering
	\subfloat[\ion{He}{II} blueshifted: $6400 - 6412$ $\ang;$ blue interval in Fig. \ref{fig:0943-emission}(d)]{\includegraphics[width=0.5\textwidth]{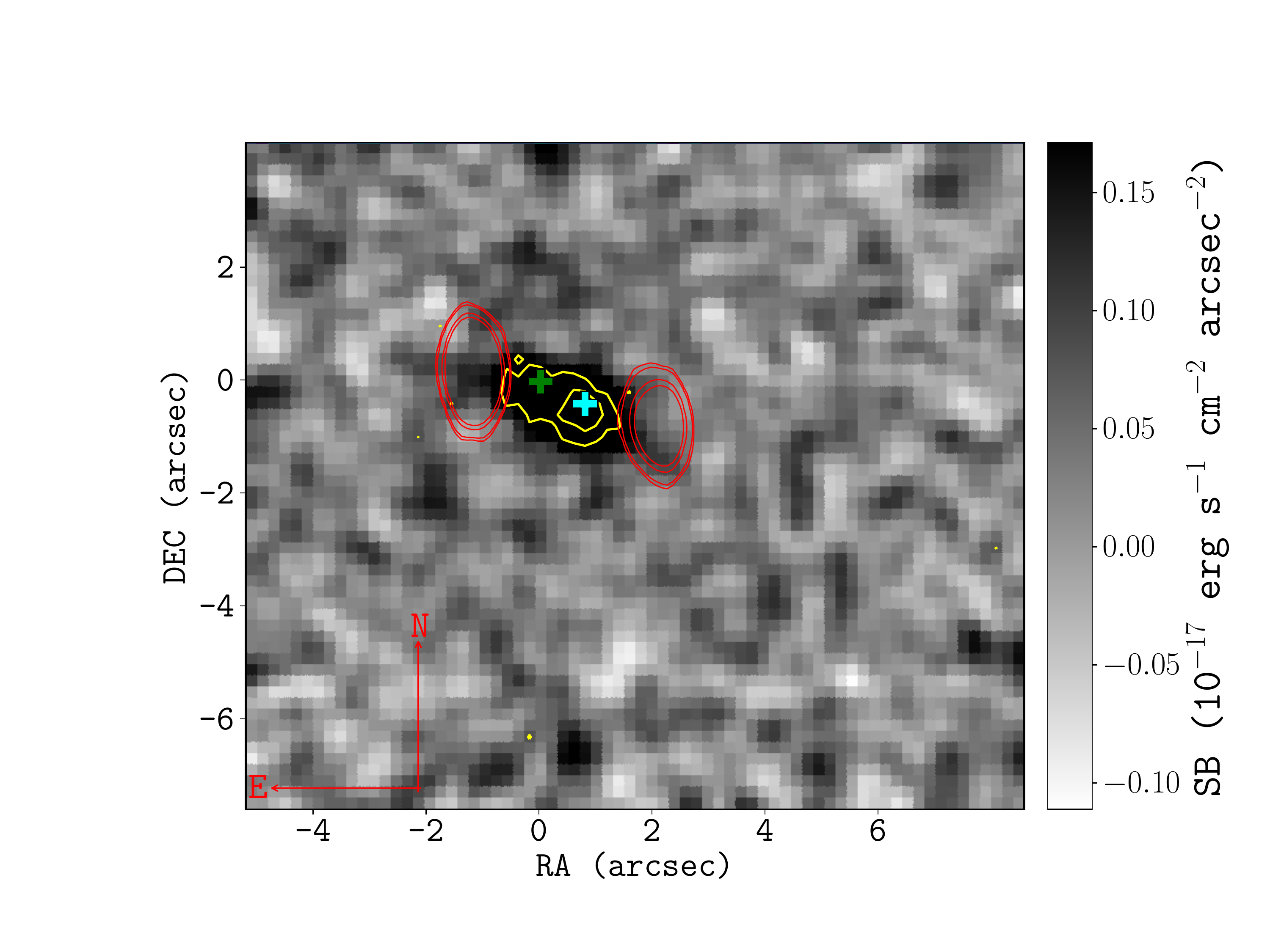}}
	\subfloat[\ion{He}{II} diffuse: $6425 - 6430$ $\ang;$ green interval in Fig. \ref{fig:0943-emission}(d)]{\includegraphics[width=0.5\textwidth]{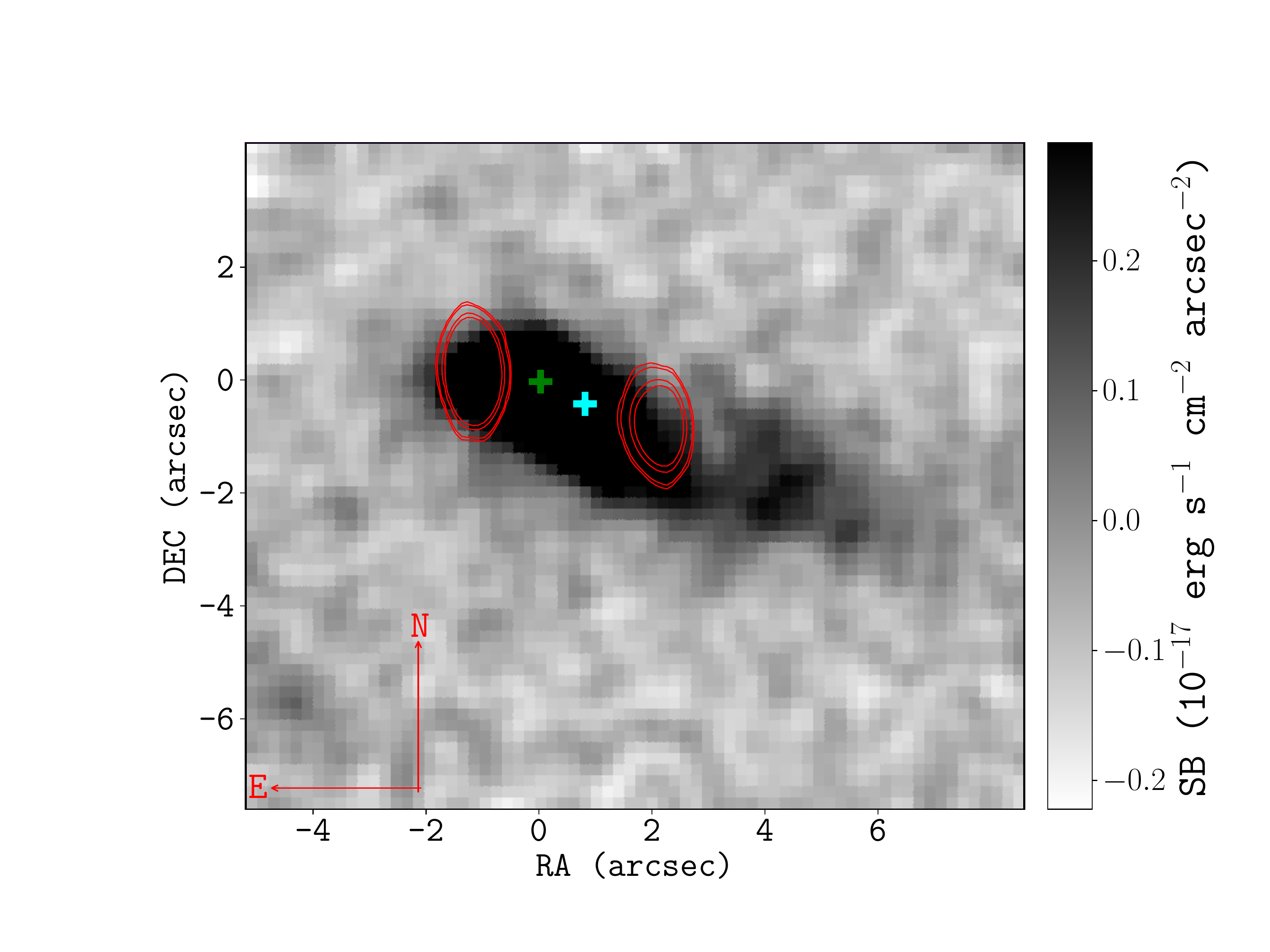}}\\
	\subfloat[\ion{He}{II} redshifted: $6445 - 6450$ $\ang;$ red interval in Fig. \ref{fig:0943-emission}(d)]{\includegraphics[width=0.5\textwidth]{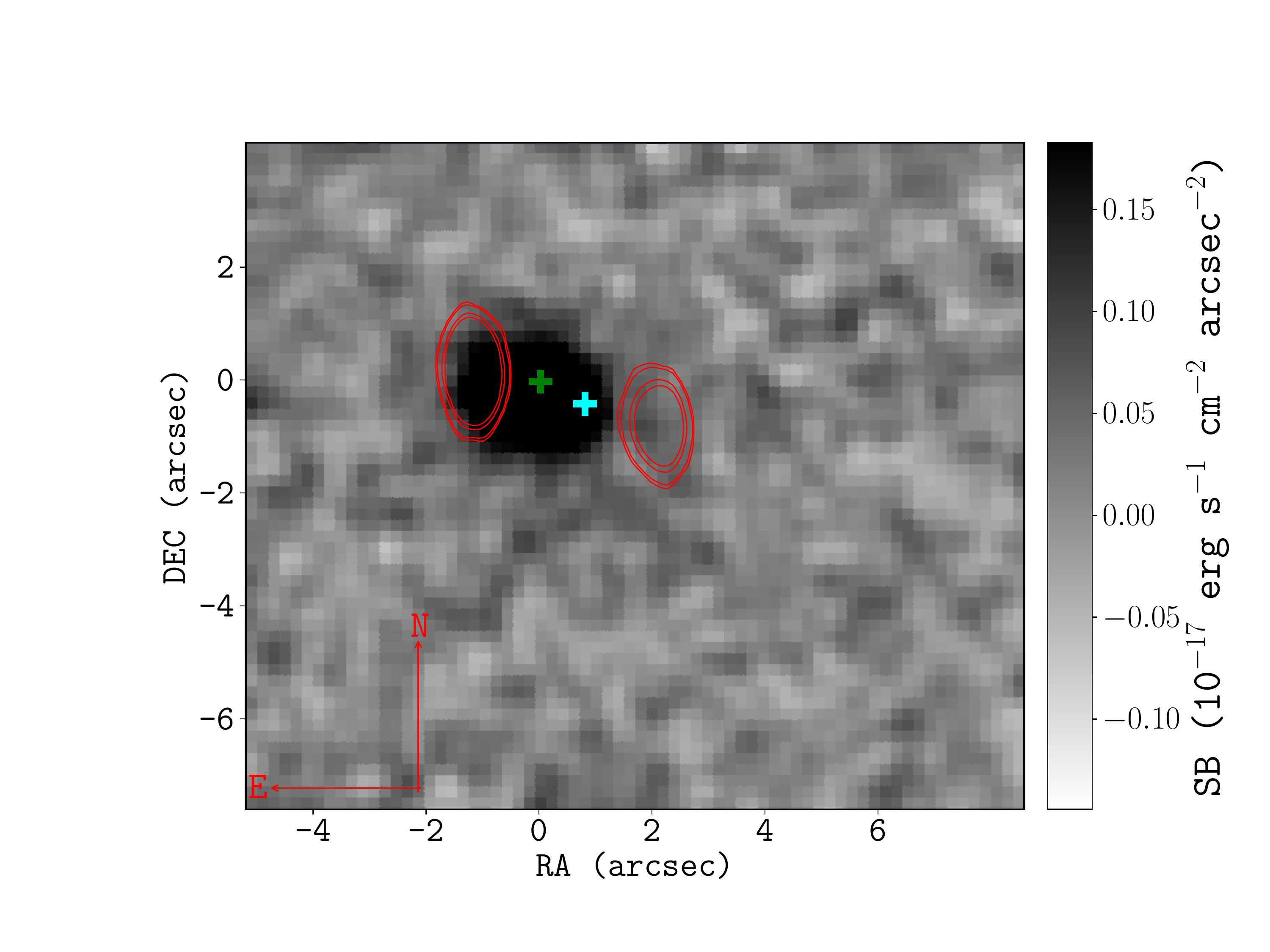}}
	\subfloat[\ion{He}{II} line emission]{\includegraphics[width=0.45\textwidth]{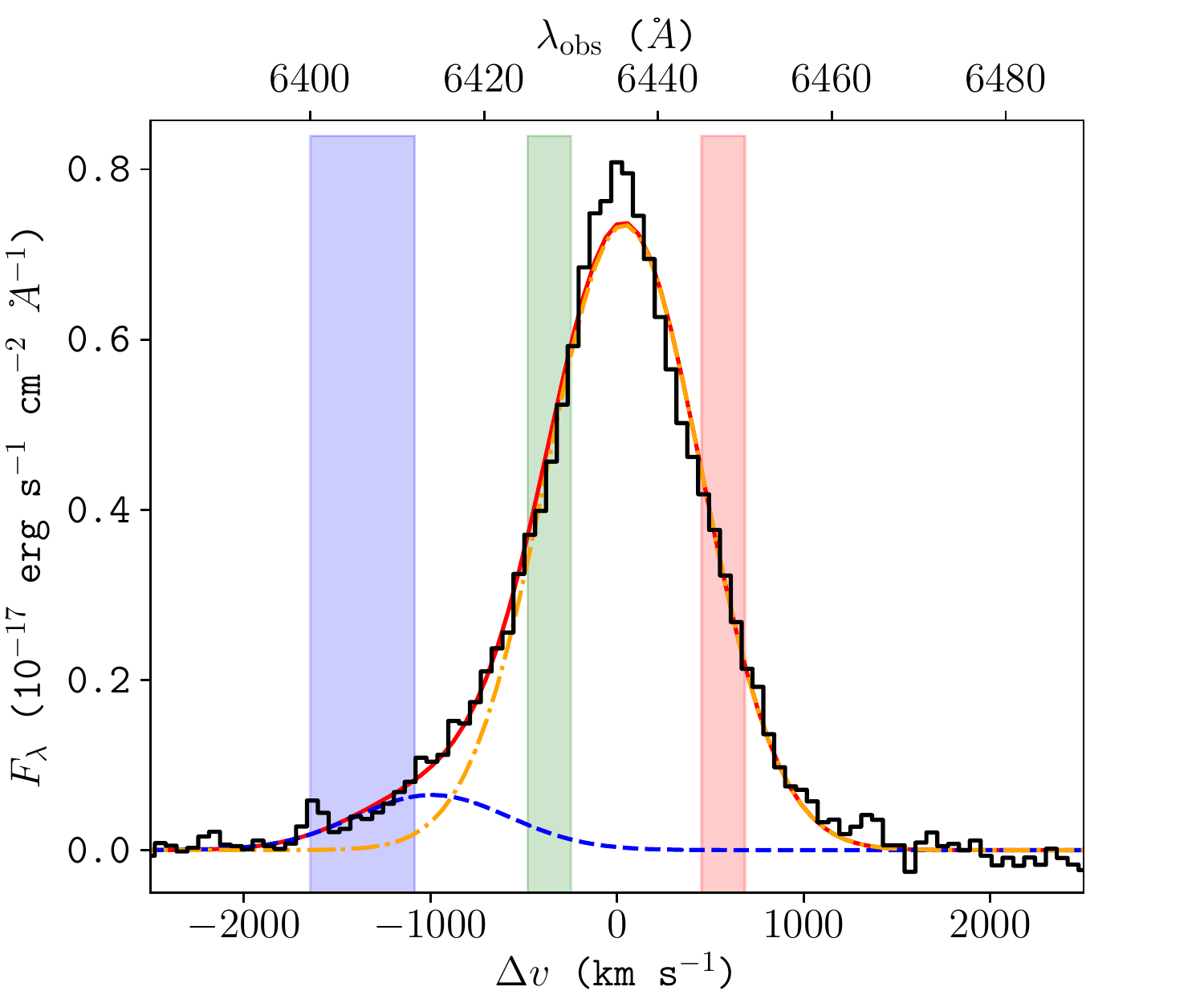}}
\caption{MUSE continuum-subtracted, narrow band images showing the spatial variation in \ion{He}{II} emission at different wavelength bands. The velocity intervals for these narrow-band images are displayed on the \ion{He}{II} profile in Fig \ref{fig:0943-emission}(d) which is a spectrum extracted from an aperture centred at the HSBR over a radius of R = 0.6\arcsec). The green and cyan crosses show the centres of the high surface brightness region (HSBR) and offset \ion{He}{II} emission from which line profiles in Fig. \ref{fig:hst-img-muse-contours} are extracted. The 4.7 GHz Very Large Array (VLA) radio contours are shown in red at the levels: $3\sigma,$ $3\sqrt{2}\sigma,$ $9\sqrt{2}\sigma,$ and $15\sqrt{2}\sigma$ for $\sigma = 2\e{-5}$ Jy beam$^{-1}$). The blueshifted component is detected over $6400 - 6412$ $\ang,$ in the blue interval, and shown as a narrow-band image in Fig. \ref{fig:0943-emission}(a). In this figure, the sub-structure of the emission is also shown using MUSE \ion{He}{II} contours with surface brightness (SB) levels: (0.038, 0.079, 0.120) $\times 10^{-17}$ erg s$^{-1}$ cm$^{-2} {\rm arcsec}^{-2}$ in yellow. The diffuse \ion{He}{II} gas, represented by the green spectral range i.e. $6425 - 6430$ $\ang,$ extends in the direction of the south-western lobe and shown in Fig. \ref{fig:0943-emission}(b). In the red spectral interval i.e. $6445 - 6450$ $\ang,$ a redshifted component is visible, in Fig. \ref{fig:0943-emission}(c), that has a clear spatial association with the north-eastern radio lobe.}
\label{fig:0943-emission}
\end{figure*}

Obtaining a narrow-band image over a smaller wavelength range of $6400 - 6412$ $\ang$ (blue interval in Fig \ref{fig:0943-emission}(d)) allows us to disentangle the blueshifted \ion{He}{II} emission from the \ion{He}{II} emission at the systemic velocity. \ion{He}{II} emission from the blueshifted component only is shown in Fig \ref{fig:0943-emission}(a) which proves that emission from the blue wing of the \ion{He}{II} line is indeed spatially offset from the high surface brightness region. 

\ion{He}{II} emission at the systemic velocity is less concentrated. In fact over the green interval in Fig. \ref{fig:0943-emission}(d), \ion{He}{II} is diffuse and elongated well beyond the radio lobes indicating possible jet-gas interactions. At the red wing of the line, \ion{He}{II} emission is concentrated at the eastern lobe and HSBR as Fig. \ref{fig:0943-emission}(c) shows. 

Both \lya and \ion{C}{IV} have evidence for turbulent blueshifted motions from the line-fitting presented (see sections \ref{section:Lya-fit} and \ref{section:CIV-NV-fit}). If the blueshifted component is indeed an outflow, this implies that the outflowing gas contains more than one ionised gas tracer which we can expect for a bulk outflow of gas from the enriched ISM. 

\begin{table*}
\caption[]{Best fit results to \ion{He}{II} high surface brightness region (HSBR) and offset lines in Fig. \ref{fig:hst-img-muse-contours}. The emission is also classified as blueshifted relative to the systemic or emitted from the systemic velocity.}    
\centering                          
\begin{tabular}{ l l D{,}{\, \,\pm\, \,}{-3} D{,}{\, \,\pm\, \,}{-3} D{,}{\, \,\pm\, \,}{-3} D{,}{\, \,\pm\, \,}{-3} D{,}{\, \,\pm\, \,}{-3} }
\hline\hline MUSE  \\
\hline            
\ion{He}{II} line region 		& 
Component &  
\mc{Line centre (rest)} & 
\mc{Line centre (obs.)} 
&\mc{Line flux}  
& \mc{Line width}
& \mc{Velocity}   \\ 
& 
&  
\mc{$\lam_0$ (\ang)}  
& \mc{$\lam$ (\ang)} 
& \mc{$F$ (10$^{-17}$ erg s$^{-1}$ cm$^{-2}$)}
& \mc{FWHM (km s$^{-1}$)} 
& \mc{$\Delta \varv$} \\  
&  &  \mc{} & \mc{} &\mc{} \\  
\hline     
&  &  \mc{} & \mc{} &\mc{} \\  
HSBR 	& blueshifted 	& 1640.40 & 6413.49,3.04 	& 1.43,0.47 		& 970, 225 	& -1053,3206  \\  
								& systemic   		& 	\ditto		& 6436.09,0.30 	& 16.42,4.86 	& 978, 24	 	& 0.02,0.01 \\ 
&  &  \mc{} & \mc{} &\mc{} \\  			  	  	
Offset 	& blueshifted 	& \ditto 		& 6414.18,3.21 	& 2.69,0.81 		& 1134, 195 	& -1025,3251\\  
								& systemic   		& 	\ditto		& 6434.76,0.83 	& 7.64,0.80 		& 993,48 		& -46,27 \\   
&  &  \mc{} & \mc{} &\mc{} \\   
\hline                                   
\end{tabular} 
\label{table:HeII-nucleus-offset}  
\end{table*}

\section{Quasar and HzRG absorbers, in comparison}\label{section:hzrg-vs-quasar-abs}
The absorption lines in a quasar continuum emerge from absorption by intervening gas in the inter-galactic medium and the circumgalactic mediums of foreground galaxies \citep[e.g.,][]{bechtold2001}. In this work, we use quasar absorption lines to determine whether the absorbers in Yggdrasil are associated with the halo of the galaxy or the IGM, assuming that HzRG and quasar absorbers are drawn from the same parent population. 

\begin{figure*} 
\captionsetup[subfigure]{labelformat=empty}
\hspace{1pt}
\centering
\subfloat[]{\includegraphics[width=0.5\textwidth]{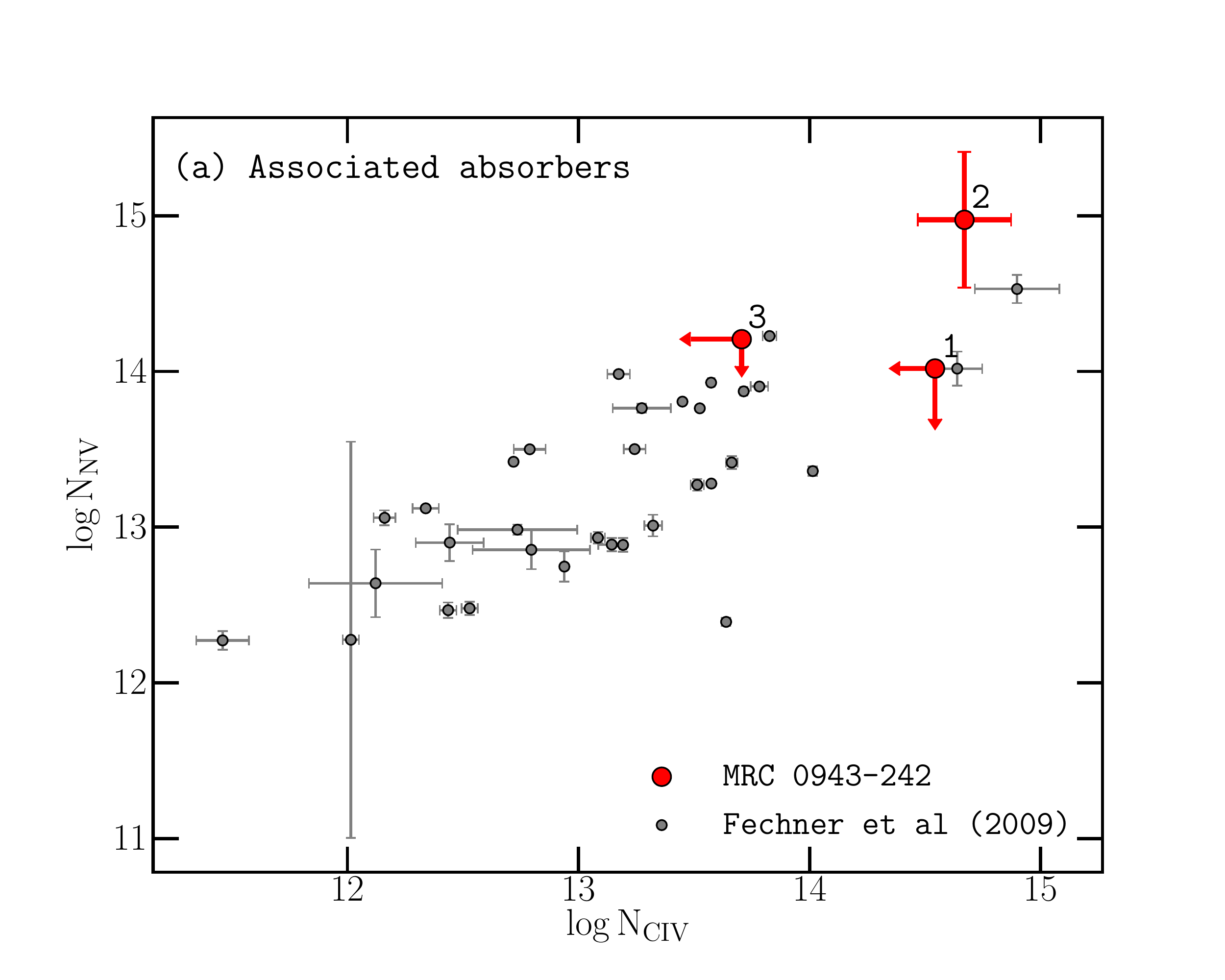}}
\centering
\subfloat[]{\includegraphics[width=0.5\textwidth]{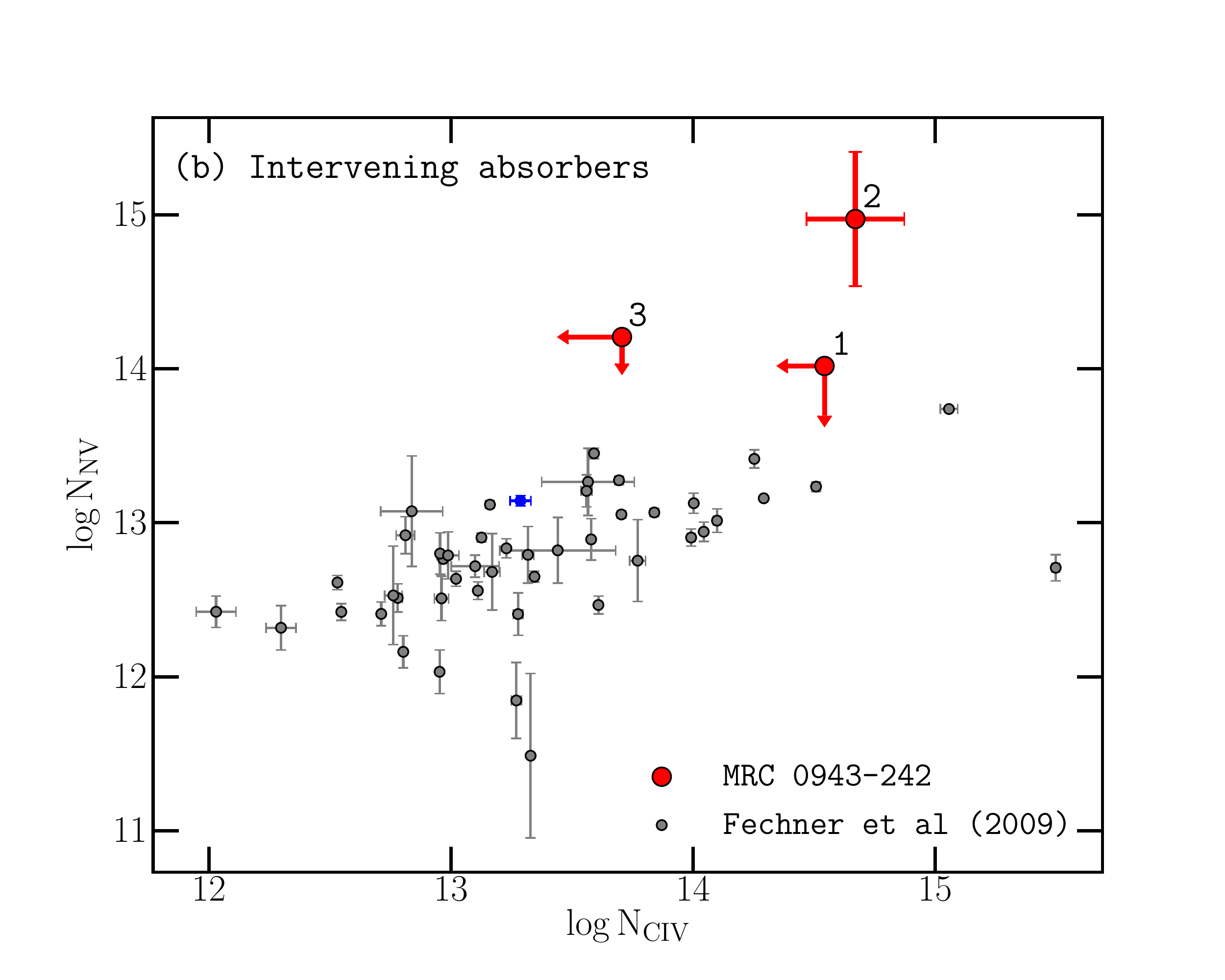}}
\caption{\ion{N}{V} and \ion{C}{IV} column densities of associated ($left$) and intervening ($right$) absorbers from \citet{fechner2009} (in grey and blue: secondary detections for the quasar) and Yggdrasil detections (red).}
 \label{fig:NV-quasar-absorption}
\end{figure*}

To do this, we compare our results to quasar absorption parameters of \ion{H}{I}, \ion{N}{V} and \ion{C}{IV} for associated quasar absorbers from \citet{fechner2009}. In this work, the absorbing gas identified as intervening is located at velocity shifts of $\lvert \Delta \varv \rvert > 5000$ km s$^{-1}$ from the quasars in their sample. The  associated absorbers are those at velocity shifts, $\lvert \Delta \varv \rvert > 5000$ km s$^{-1}$ from the quasars. We show this comparison in Fig \ref{fig:NV-quasar-absorption}. Our results (which generally have velocity shifts, $\lvert \Delta \varv \rvert < 1500$ km s$^{-1}$ for all absorbers) are in better agreement with absorption by associated gas rather than intervening. This implies that the absorbers in Yggdrasil are more likely to be associated with the galaxy i.e. the absorbers are bound to the ISM and/or CGM, if they are similar or from the same parent distribution.

\begin{figure} 
\centering
\includegraphics[width=\columnwidth]{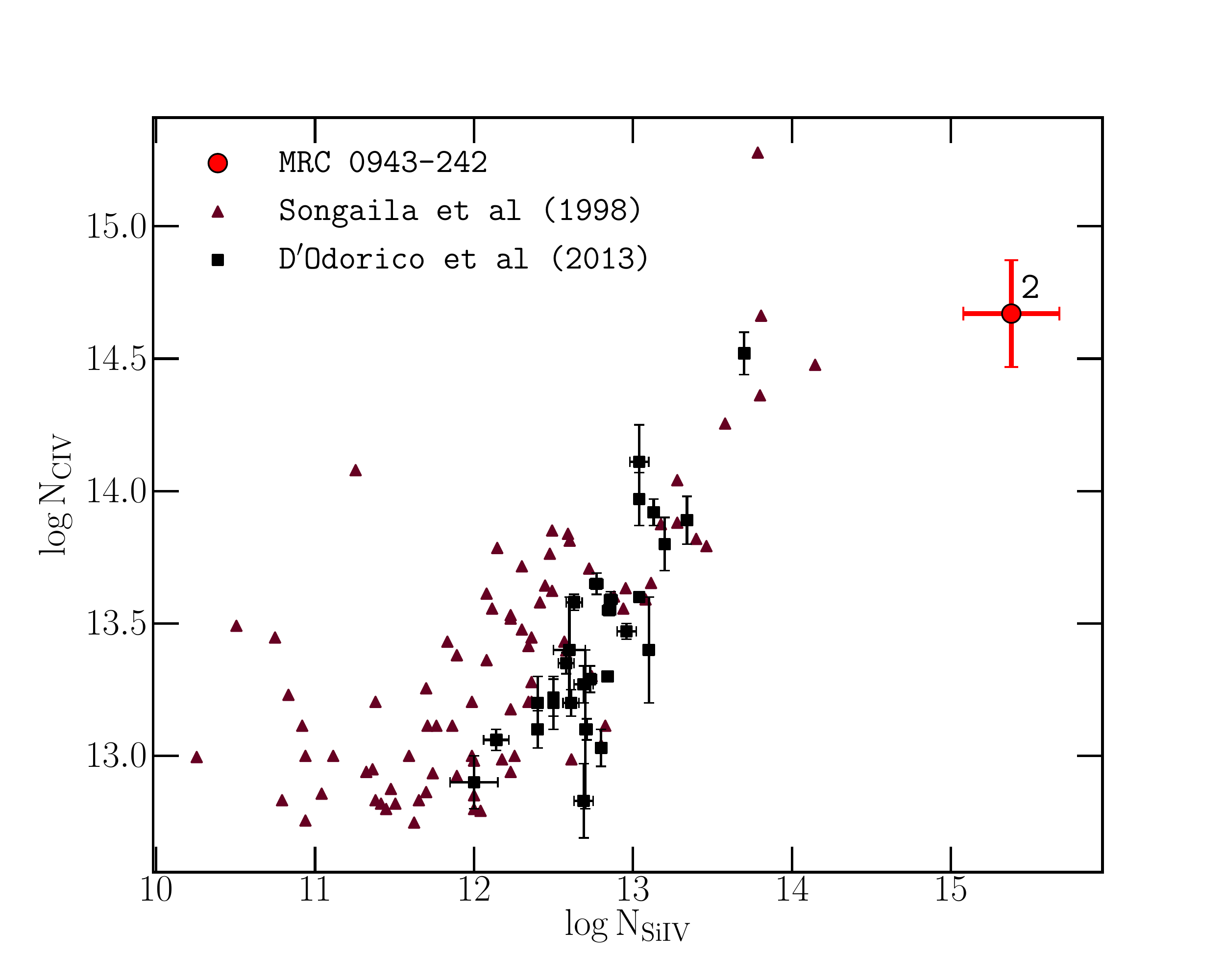}
\caption{$N_\ion{C}{IV}$ relative to $N_\ion{Si}{IV}$ for absorbers in the IGM in \citet{songaila1998} and quasar associated absorbers in \citet{dodorico2013}.}
\label{fig:SiIV-quasar-absorption}
\end{figure}

We have an additional set of quasar absorbers with which to compare the \ion{C}{IV} and \ion{Si}{IV} column densities measured. Quasar absorption for the ionised gas, in particular, have been measured by \citet{songaila1998} and \citet{dodorico2013}. In the former, absorption is from the IGM while the latter results show quasar associated absorption at much higher redshifts i.e. z $>$ 6. This comparison is shown for Fig. \ref{fig:SiIV-quasar-absorption}. As seen from Fig. \ref{fig:SiIV-quasar-absorption}, Yggdrasil absorbers are in better agreement with associated absorbers of \citet{dodorico2013} than the intervening absorbers of \citet{songaila1998}. These order of magnitude comparisons show that the absorbers in the Yggdrasil spectrum are much more likely to be associated with the galaxy halo.

\section{Photoionisation Modelling}\label{section:photoionisation-modelling}
Using photoionisation modelling code to determine the main ionising mechanisms producing the observed ionised gas and its column densities in the absorbing gas shells surrounding Yggdrasil. A very detailed study of this has been carried out in \citet{binette2000} who obtain a metallicity of Z/Z$_\odot \simeq 0.02$ (Z$_\odot$ being the Solar metallicity) for the strongest of the \lya absorbers. Their findings also indicated that emitting and absorbing gas are not co-spatial and that the gas is best described by an \ion{H}{I} volume density of n$_\ion{H}{I} = 100$ cm$^{-3}.$ In addition to this, \citet{binette2006} sought to determine the ionising mechanism behind the absorbers in this source, and thus showed that stellar photoionisation results in the \ion{H}{I} and \ion{C}{IV} absorber column densities 

It is worth noting that their conclusions were obtained using only column densities for \ion{H}{I} and \ion{C}{IV}. With our recent detection of \ion{N}{V} absorption at the same velocity as the the strong \lya absorber, we can provide more insight to these findings. Given the detection of \ion{N}{V}, flux from the metagalactic background may be too weak to ionise the absorbers. At z $\geq$ 2.7, \ion{He}{II} reionisation was more active than it is at lower redshifts meaning that ionising photons with energies of $E_\gamma > 54.4$ eV had Mpc-scale mean free paths \citep{shull2010,mcquinn2016}. The metagalactic background is therefore less likely to be a major ionising mechanism powering the absorbers in the halo of Yggdrasil. 

We also consider the possibility that shocks by the powerful radio jets may ionise the gas. This was proposed by \citet{dopita1995}, and observational evidence for shock ionisation in HzRGs has been shown by several authors, particularly in sources with relatively small radio sources still contained within the host galaxy \citep[e.g.][]{allen1998,best2000,debreuck2000}. A powerful diagnostic of shock ionised emission are the line ratios of carbon: in shock ionised regions, we expect the \ion{C}{II]} $\lam2326$ line to be as strong as its higher ionisation counterparts. Our MUSE spectrum (Fig. \ref{fig:0943-spectrum}) contains three carbon lines, and allows us to make this test. Interestingly, the \ion{C}{II]} $\lam1906,1908$ line contains a much stronger blueshifted component than the \ion{C}{III]} and \ion{C}{IV} $\lam\lam1548,1550$ lines. This high \ion{C}{II]}/\ion{C}{III]} and \ion{C}{II]}/\ion{C}{IV} ratios are consistent with shock ionisation \citep[see e.g. Fig. 1 of][]{best2000}. This lends evidence to our claim that this blueshifted gas near the western radio lobe is an outflow driven by the radio jet. On the other hand, the radio of the main component near the AGN has a \ion{C}{III]}/\ion{C}{II]} ratio of $\sim$2.4, which places it squarely in the AGN photo-ionisation region. As this main velocity component dominates the flux of all the other emission lines, we therefore do not consider shock ionisation as a relevant contributor in those lines.

Stellar photoionisation is a possibility as well. The star-formation rate (SFR) in Yggdrasil, however, is SFR=41 M$_\odot$ yr$^{-1}$ \citep{falkendal2019} and may be insufficient in producing the star-burst activity that would ionise the absorbers. Another possibility is the AGN, which we can expect to produce a strong enough ionising continuum to produce the detected absorption. 

We, therefore, revisit the analysis by running photoionisation models using \pkg{cloudy c17} \citep{ferland2017} --  a spectral synthesis code designed to run grid models that provide simulated chemical abundances in photoionised gas. Within the halo of Yggdrasil, we have shown that \lya absorbers are at the same velocity as \ion{C}{IV}, \ion{N}{V} and \ion{Si}{IV} absorbers suggesting that the gas tracers occupy the same gas volume. 

The ionisation parameter is defined as $U(r) = \frac{ 1 }{ 4 \pi {\rm r}^2 {\rm c} {\rm n}_\ion{H}{I}} \int^\infty_{\nu_0} \frac{ {\rm L}_\nu }{ h\nu }d\nu,$ is the quotient of the volume density of hydrogen ionising photons and the volume density of hydrogen atoms contained within the gas. Here, $r$ is the line-of-sight distance from the ionising source to the gas, n$_\ion{H}{I}$ is the hydrogen density, $L_\nu$ is the specific luminosity, and $\nu$ the observing frequency such that the integral, $\int^{\infty}_{\nu_0} \frac{ {\rm L}_\nu }{ h\nu }d\nu,$ is the luminosity of the ionising source. Provided that we can determine $U(r)$ for the radiation field ionising the absorbing gas, we can also compute the distance of the gas from the radiation source  along the line-of-sight.

\subsection{AGN Photoionisation of the Absorbers}\label{section:photoionisation-absorbers}

We have shown, using a comparison to quasar absorbers in \citet{fechner2009}, that those in Yggdrasil are located within the CGM rather than IGM. In the \pkg{cloudy} model, we use a power-law (PL) spectral-energy distribution \pkg{SED} (where the ionising flux is given by S$_\nu \propto \nu^\alpha$) to simulate the incident radiation produced by the AGN. In this, we tested both $\alpha=-1.5$ and $\alpha=-1.0$ to model photoionisation of the gas by an AGN \citep[e.g.,][]{villar-martin1997,feltre2016,humphrey2018}.

The model grid consists of a set of input parameters. The hydrogen density is fixed to n$_\ion{H}{I} = 100 {\rm cm}^{-3}$ which is commonly adopted to describe low density, diffuse gas which we expect in the extended absorbers \citep[e.g.,][]{binette2000,humphrey2008a}. The metallicity (Z) is varied over a range of values between Z/Z$_\odot$ = 0.01 and 10. The \ion{H}{I} column density N$_\ion{H}{I},$ is taken from the line-fitting results (shown in Table \ref{table:absorption-fits}). The ionisation parameter varies over the range, $ 10^{-2.5} < U < 10^{-1}$ in 0.5 dex increments.

We do not include the effects of dust extinction because the surface brightness of emission re-radiated by dust, detected by ALMA (the Atacama Large Millimetre/sub-mm Array) at an observing frequency of $\nu_{\rm obs}$ = 235 GHz, is comparatively low in Yggdrasil compared to the offset region at projected distances of $r = 65$ to 80 kpc from the host galaxy where three resolved dusty companions are observed (G16). This observation implies that dust has a negligible contribution to the \pkg{SED} at the host galaxy. Moreover, \citet{falkendal2019} have shown that the star-formation rate (SFR) in Yggdrasil, is low in comparison to that which is detected in the companion sources, collectively. 

In the \pkg{cloudy} models, we have assumed an open or plane-parallel geometry for the absorbing gas medium which implies that the cloud depth is significantly smaller than its inner radius or, more simply, the distance between the ionising continuum source and the illuminated gas surface. Following the findings from \citet{binette2000}, we assume that the absorbers are not co-spatial with the emitting gas. 

\begin{figure}
\centering 
\includegraphics[width=\columnwidth]{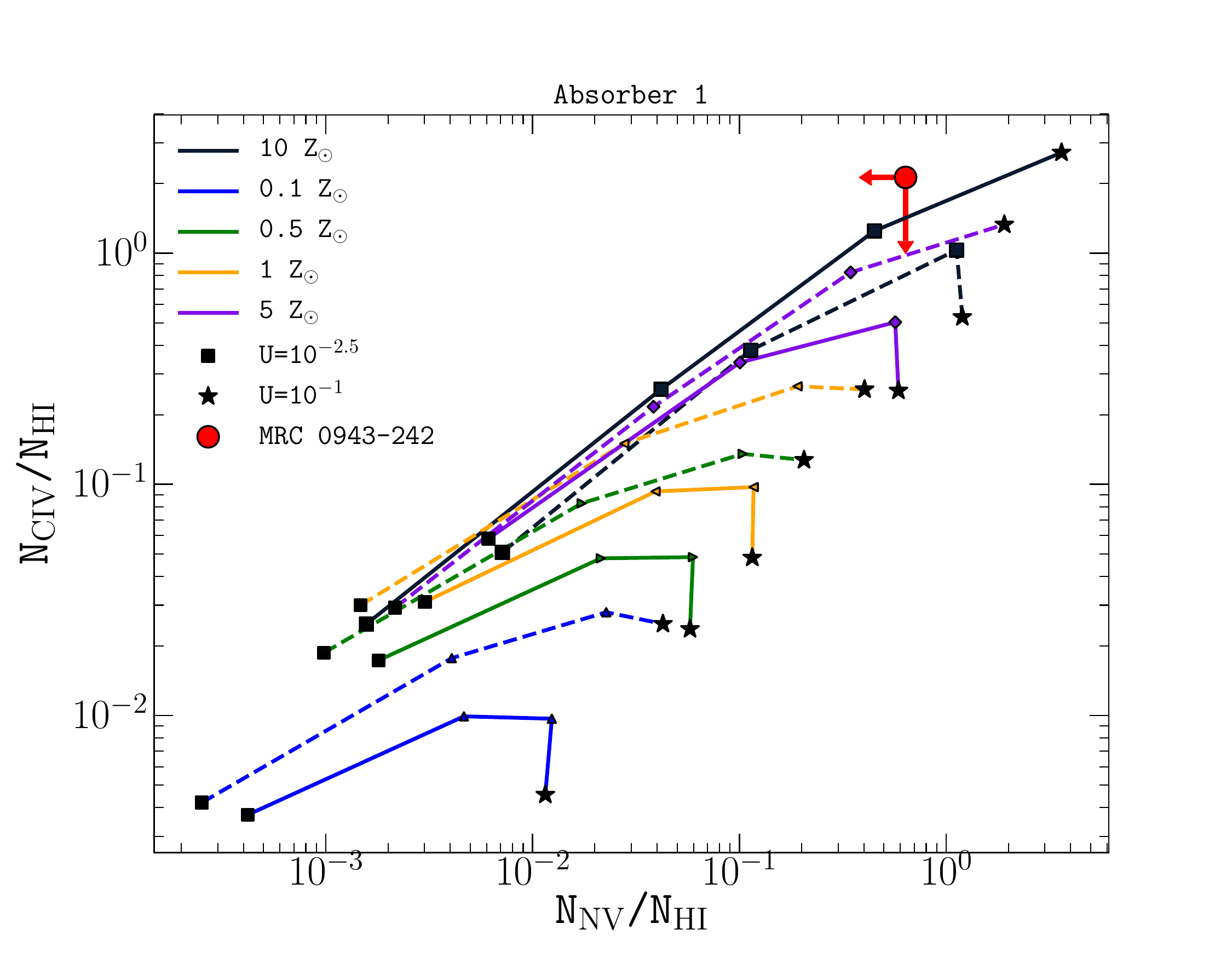}
\caption{\ion{N}{V} and \ion{C}{IV} column densities obtained from \pkg{cloudy} photoionisation models. The predicted column densities are for absorber 1 which has an \ion{H}{I} column density of $N_{\ion{H}{I}} = 10^{14.2}$ cm$^{-2}$ that is kept constant as $U(r)$ increases from $U=10^{-2.5}$ to $U=10^{-1}.$ in increments of 0.5 dex. The metallicities, Z/Z$_\odot$ = 0.1, 0.5, 1 and 5 are tested for spectral indices $\alpha=-1.0$ (solid lines) and $\alpha=-1.5$ (dashed lines).}
\label{fig:cloudy-abs1}
\end{figure}

\begin{figure}
\centering
\includegraphics[width=\columnwidth]{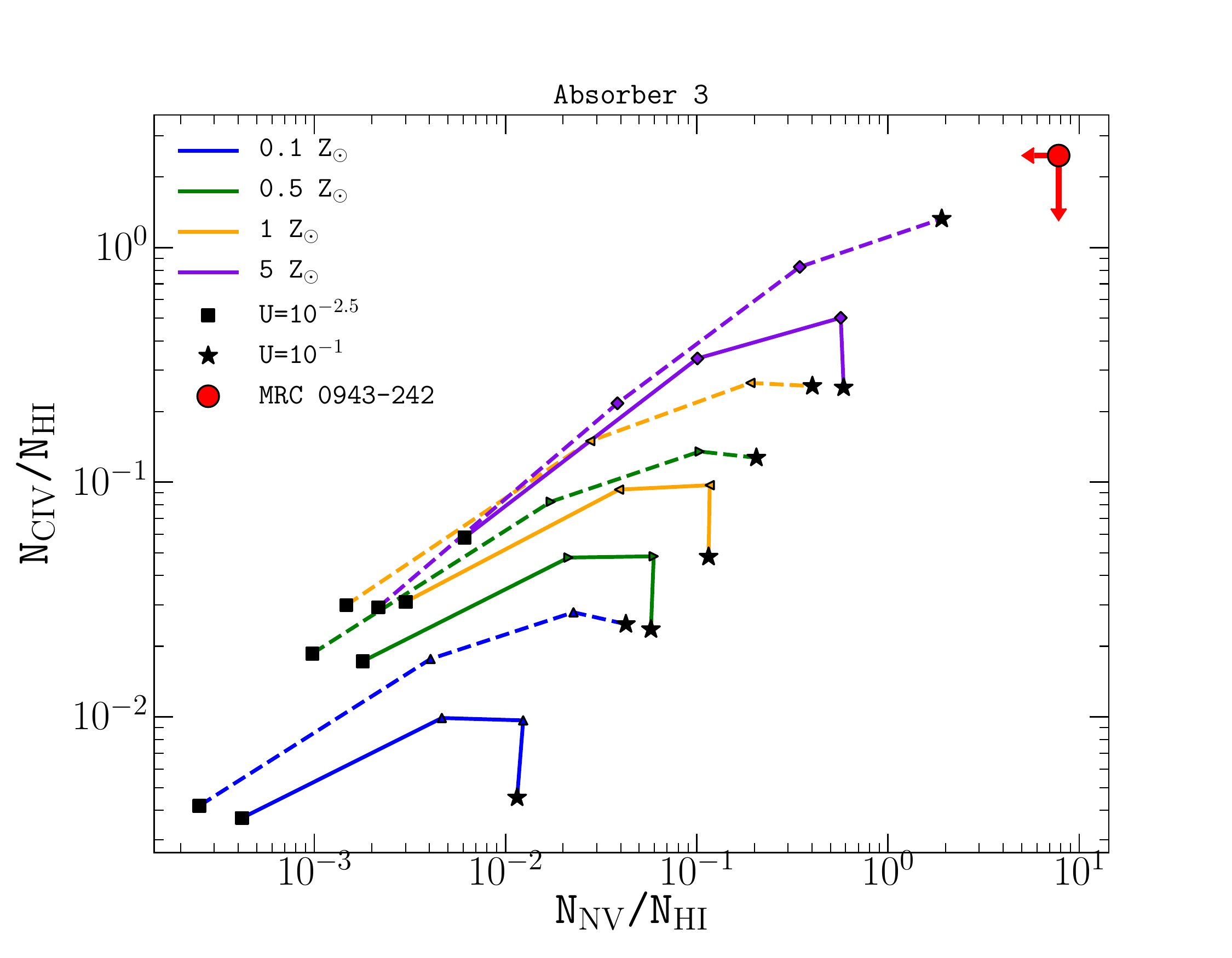}
\caption{\ion{N}{V} and \ion{C}{IV} column densities obtained from \pkg{cloudy} photoionisation models similar to those in Fig \ref{fig:cloudy-abs1}. The models shown here are for absorber 3 since the \ion{H}{I} column density is kept constant at $N_{\ion{H}{I}} = 10^{13.3}$ cm$^{-2}.$ The metallicities tested are Z/Z$_\odot$ = 0.1, 0.5, 1 and 5.}
\label{fig:cloudy-abs3}
\end{figure}

\begin{figure}
\centering
\includegraphics[width=\columnwidth]{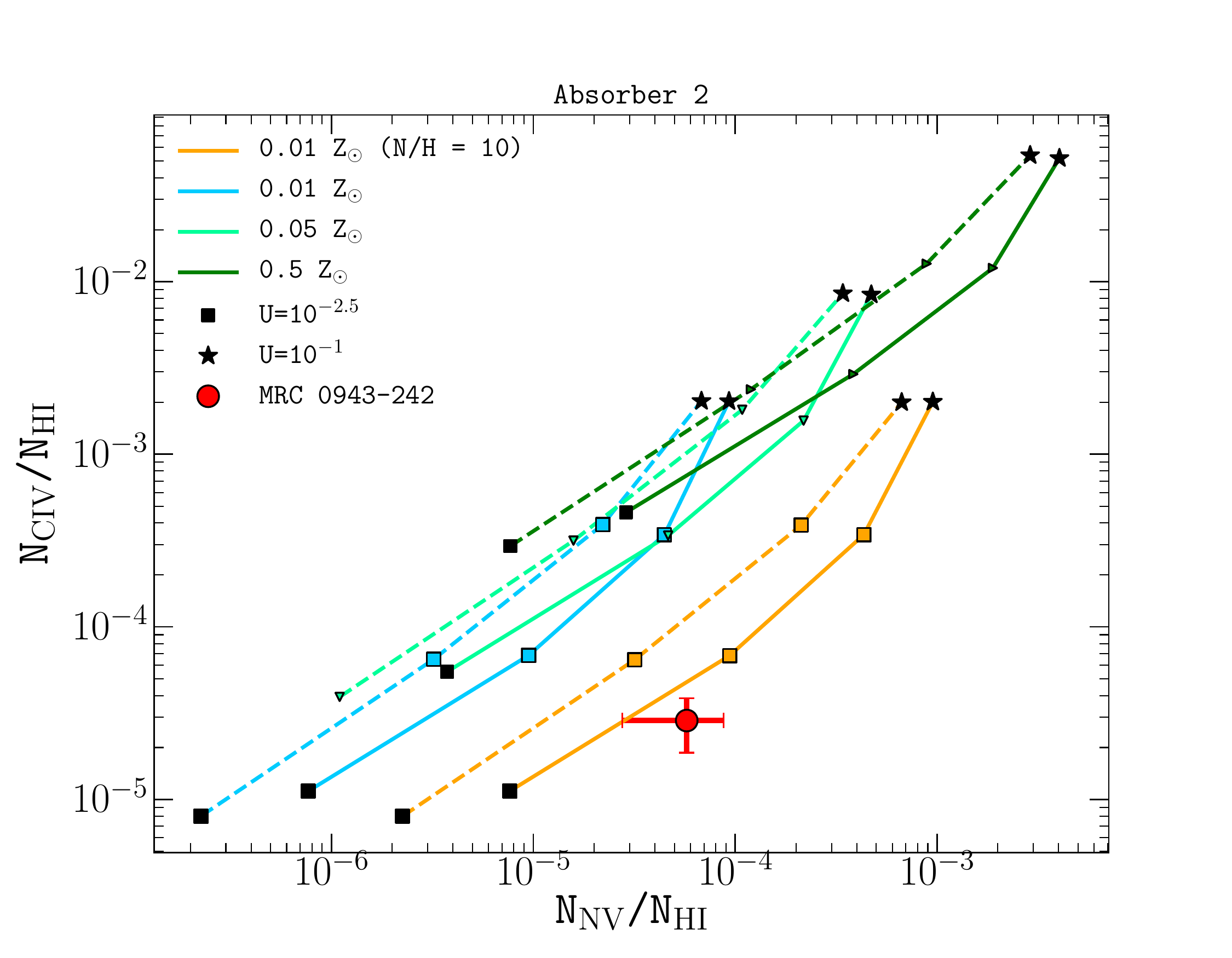}
\caption{\ion{N}{V} and \ion{C}{IV} column densities obtained from photoionisation \pkg{cloudy} models similar to those in Fig. \ref{fig:cloudy-abs1}. The models shown here are for absorber 2, the strong absorber, since the \ion{H}{I} column density is kept constant at $N_{\ion{H}{I}}$ = $10^{19.2}$ cm$^{-2}.$ The nitrogen abundance at Z/Z$_\odot = 0.01$ has also been enhanced by a factor of 10 (orange curve) i.e. N/H = 10.}
\label{fig:cloudy-abs2}
\end{figure}

The column density measures we obtain are shown against the simulated column densities in Figs \ref{fig:cloudy-abs1} and \ref{fig:cloudy-abs3} and the results indicate that absorbers 1 and 3 may have supersolar metallicities of Z/Z$_\odot \simeq$ 10 and 5, respectively. However, due to uncertainties in line-fitting, the column densities obtained are upper limits and no proper conclusion can be drawn about the metallicities of the absorbers. If these models do suggest super-solar metallicity in absorbers. this would not be a unusual given that super-solar metallicites in HzRG absorption line gas have been measured before \citet{jarvis2003,binette2006}. In \citet{jarvis2003}, absorbers in the halo of the z=2.23 radio galaxy, 0200+015, are shown to have Z/Z$_\odot \sim$ 10 as well as covering factors of C $<$ 1.0. This implied that the absorbers are likely to be co-spatial with the extended emission line region. Based on Fig. \ref{fig:cloudy-abs2}, the measured column densities of absorber 2 prove that its metallicity is closer to a value of Z/Z$_\odot = 0.01.$  The results also suggest that either the \ion{N}{V} column density is under-predicted by the Z/Z$_\odot = 0.01$ model. We explore the possible reasons for this in the following section. 

\subsection{Nitrogen abundances under-predicted by AGN photoionisation}

The assumed power laws for the ionising continuum do not re-produce the metal ion column densities observed in the strong absorber (absorber 2). We are certain that the \ion{C}{IV} column density is well measured given its agreement to literature values (\citealp{jarvis2003}; G16). The \ion{N}{V} cannot be compared to any previous results other than the quasar absorbers shown in Fig. \ref{fig:NV-quasar-absorption} but has been fit with a similar level of uncertainty as \ion{C}{IV}. The results shown in Fig. \ref{fig:cloudy-abs2} indicate that AGN photoionisation alone cannot produce the \ion{N}{V} column densities measured. At the same time, a softer ionising continuum is also insufficient in producing a high ionisation tracer such as \ion{N}{V}. In other words, a power law under-predicts the measured \ion{N}{V} column density for the strong \lya absorber. We, therefore, explore secondary nitrogen production, enhanced nitrogen abundances and lower \ion{H}{I} column density as possible reasons for under-predicted nitrogen abundances. 

\subsubsection{Secondary nitrogen production}

The under-predicted nitrogen abundance may result from scaling the nitrogen abundance (relative to hydrogen) according to the solar value i.e. N/H $\propto$ (N/H)$_\odot.$ We propose that secondary nitrogen production by intermediate mass stars with M/M$_\odot=$ 1 - 8 \citep[e.g.,][]{henry2000}. In this process, abundances scale as N/O $\propto$ O/H $\propto$ Z (with N, O and H being the nitrogen, oxygen and hydrogen abundances and Z, the metallicity of the gas). This implies that N/H $\propto$ (O/H)$^2$ $\propto$ Z$^2.$ However, secondary nitrogen production is known to occur in in gas with super-solar metallicities (i.e., Z/Z$_\odot \geq 1.0$) while primary nitrogen production dominates at sub-solar metallicities \citep[e.g.,][]{Hamann&Ferland1993}. We can consider secondary nitrogen the cause of nitrogen enhancement in absorber 2. Fig. \ref{fig:NV-quasar-absorption}(a) shows that the absorbers in Yggdrasil have similar column densities as associated quasar absorbers in the \citet{fechner2009} sample. In this study, secondary nitrogen production is posited as a reason for the nitrogen enrichment of the quasar absorbers.  

If secondary nitrogen production from carbon and oxygen in the CNO cycle causes the enhanced nitrogen abundance in the strong absorber, the absorber would have a higher metallicity than originally determined \citep[Z/Z$_\odot \simeq 0.02$;][]{binette2000,binette2006}. As is the case for the extended emission line region (EELR) in S18 where is said to occur secondary nitrogen production occurs for a gas metallicity of Z/Z$_\odot \simeq 2.0.$ The \pkg{cloudy} model grid indicates that in order for gas to be enriched by secondary nitrogen, it would also have a metallicity of Z/Z$_\odot \geq 1.0.$ This is improbable given that an increase in the cloud metallicity results in an increase in both \ion{C}{IV} and \ion{N}{V} column densities. Hence, we exclude secondary nitrogen production as a mechanism for the enhancement of nitrogen abundances. 

\subsubsection{Degeneracy of $b$ and N$_\ion{H}{I}$ for the strong absorber}

The degeneracy between the \ion{H}{I} column density $N_\ion{H}{I}$ and the Doppler parameter have been described in S18 who found two probable best fit solutions for the $b$ parameter and N$_\ion{H}{I}$ in absorber 2. These are $N_\ion{H}{I}/{\rm cm}^{-2} = 10^{19.63}$ with $b = 52$ km s$^{-1}$ and $N_\ion{H}{I}/{\rm cm}^{-2} = 10^{15.20}$ with $b = 153$ km s$^{-1}.$ Taking the lower \ion{H}{I} column density solution as well as the $N_\ion{C}{IV}$ and $N_\ion{N}{V}$ measurements from our results, we ran the \pkg{CLOUDY} model grid using the same AGN photoionisation scheme (described in section \ref{section:photoionisation-absorbers}).

This time, applying the low \ion{H}{I} column density solution (i.e., $N_\ion{H}{I}/{\rm cm}^{-2} = 10^{15.20}$) to the grid resulted in the observed column density ratios (i.e., $N_\ion{C}{IV}/N_\ion{H}{I}$ and $N_\ion{N}{V}/N_\ion{H}{I}$) suggesting that absorber 2 has a supersolar metallicity, Z/Z$_\odot \simeq 5.$ This result veers greatly from the expected abundances from \citet{binette2000} i.e. Z/Z$_\odot \simeq 0.02$ which was obtained through photoionisation modelling in \pkg{MAPPINGS IC} \citep{binette1985,ferruit1997}. What this result may imply is that if the cloud had a lower \ion{H}{I} column, its metallicity would need to be higher than Z/Z$_\odot \simeq 0.01.$ 

\subsubsection{Nitrogen abundance enhancement}

For the \pkg{cloudy} models to reproduce the observed column densities, we require a nitrogen abundance that is greater than the solar abundance by a factor of 10 as Fig. \ref{fig:cloudy-abs2} suggests. We can look at occurrences of enhanced nitrogen abundances in similar sources to attempt to find an answer for this. 


The photoionisation models for shown in Fig. 10 of \citet{binette2006} seem to suggest that the \ion{N}{V} column density of absorber 2 suggests that it may be ionised by the same mechanism ionising the Lynx arc nebulae which is a metal-poor \ion{H}{II} galaxy at z $\simeq$ 3.32 that has undergone a recent star-burst \citep{villar-martin2004}. However, this scenario would require stellar photoionisation rather than AGN. The presence of \ion{N}{V} in the absorber presents a strong case for AGN rather than stellar photoionisation since it has such a high ionisation energy. 

To determine whether the current episode of AGN activity has ionised the absorber, we determine its life-span. The radio size of Yggdrasil is $r \simeq$ 29 kpc \citep{pentericci1999} and its approximate expansion speed, $varv \simeq$ 0.05c. Hence, the age of the radio source is approximately, $\tau_{\rm jet} \simeq$ 2 Myr. The strong absorber, on the other hand, is roughly, $r \simeq$ 60 kpc, and has an outflow rate of 400 km s$^{-1}$ relative to the galactic nucleus. This translates to an approximate outflow time-scale of $\tau_{\rm abs.} \simeq 200$ Myr. The radio source being a factor of 100 younger than the absorber implies that the current radio jet duty cycle can be responsible for the ionisation of the absorber however it did not eject the gas that formed absorber 2. 

To explain the chemical abundances, we propose that the absorber has been enriched by an early episode of star-formation. Metal-rich gas from the ISM is driven by star-bursts and through powerful winds, dilutes the low metallicity gas in absorber 2. Evidence of this is seen in local massive ellipticals, considered the progeny of HzRGs, where nitrogen abundances are enhanced well above that of carbon i.e. [N/Fe] $\sim 0.8-1$ and [C/Fe] $\sim$ 0.3 (for nitrogen, N, carbon, C and iron, Fe abundances) as a result of star-burst driven superwinds \citep{greene2013,greene2015}. 

\subsection{Distances of absorbers from the ionising source}

\begin{figure*} 
\centering
\includegraphics[width=0.6\textwidth]{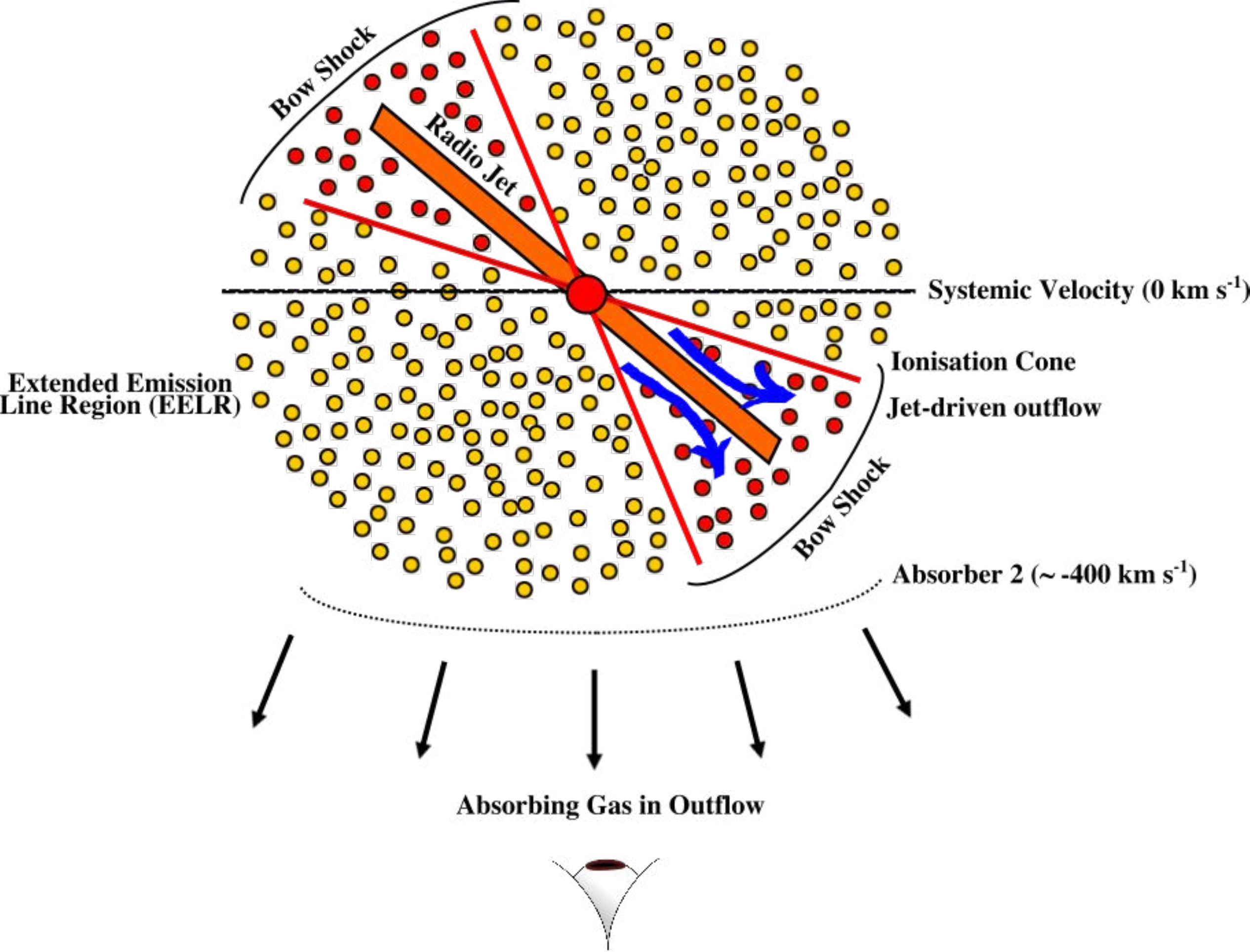}
\caption{A cross-section schematic of the circumgalactic medium of Yggdrasil based on evidence from the MUSE data and previous measures is shown. The gas has no defined spatial boundaries since its estimates size of d $\gtrsim$ 60 kpc is merely a minimum. The strong absorber screens emission originating from the T $\simeq10^4-10^6$ K extended emission line region (EELR) gas (in yellow) comprising an extended region of ionised gas as well as gas directly ionised by the AGN ionisation cone (in red). The absorbing gas (shown as dotted, curved lines) is also metal enriched due to detections of \ion{C}{IV}, \ion{N}{V} and \ion{Si}{IV} at the same velocities as \lya absorption. The blueshifted component detected in \lya and \ion{He}{II} is a jet induced outflow (blue arrows). This outflowing, ionised and turbulent gas is spatially offset from the HSBR of the halo which implies that the radio jet is tilted relative to the projected plane. Note that the locations of absorber 2 is a line-of-sight projected distance which is dependent on the ionisation parameter, $U(r).$}
\label{fig:absorption-cartoon}
\end{figure*}

The distance between the AGN and the absorbing gas can be estimated but is dependent on n$_{\ion{H}{I}}.$ Given that we cannot constrain the hydrogen density, we have only the ratio of ionisation parameters to infer the distance ratios of two absorbers e.g. $r_2 / r_1 = \sqrt{U_1 / U_2}.$ Assuming that our measure of n$_\ion{H}{I} = 100$ cm$^{-3}$ is correct, we can approximate the distance of absorber 2 from the AGN using the integrated infrared luminosity, $L_{\rm IR}^{\rm AGN},$ of the host galaxy computed by \citet{falkendal2019}. From Fig. \ref{fig:cloudy-abs2}, the ionisation parameter for absorber 2 is approximately $U \simeq 10^{-2.25}.$ Expressing the ionisation parameter as $U(r) = L/n_\ion{H}{I}r^2,$ \citep[e.g.,][]{rozanska2014}, we obtain a distance of $d \simeq 68^{+17}_{-14}$ kpc between absorber 2 and the AGN.


\section{Discussion}\label{section:discussion}
\subsection{Blueshifted emission: companion galaxy or outflow?}\label{discussion:blueshifted-emission}

There is sufficient evidence to suggest that the preferred environments of radio galaxies are rich clusters and groups. This is at low redshifts (z $\lesssim$ 1) where the most radio-loud sources have a higher likelihood of residing in kpc-scale overdense structures \citep[e.g.,][]{best2007,karouzos2014,magliocchetti2018,kolwa2019}. Radio-loud sources are also prevalent in over-dense environments consisting of dusty star-burst galaxies and/or \lya emitters at z $>$ 1.2 \citep[e.g.,][]{hatch2011,wylezalek2013,dannerbauer2014,saito2015}. Given these findings, blueshifted emission in this source systemic redshift may be from a proto-galaxy or dwarf satellite.

A well-studied example is a Minkowski's object, associated with the radio galaxy PKS 0123-016, which has undergone an increase star-formation rate due to the passage of the radio source's jets \citep{vanbreugel1985}. The high velocity dispersions of the blueshifted emission in Yggdrasil imply a different origin of the gas. Only an AGN would produce the turbulent gas motions that can broaden lines such as \ion{He}{II} to line widths of FWHM=1500 km s$^{-1}.$ Furthermore, dwarf galaxies tend to have outflow velocities of ionised gas of $\varv_{\rm out} \lesssim 100$ km s$^{-1}$ \citep{martin2005}. Such velocities are well below the outflow rates we observe in the blueshifted, ionised gas which are $\varv_{\rm out} \gtrsim 400$ km s$^{-1}$. 

The gas turbulence being driven by the jets is more likely and is a known occurrence within the gas haloes of powerful radio galaxies \citep[e.g.,][]{humphrey2006,morais2017,nesvadba2017a}. The jet-driven outflow scenario is well supported 4.7 GHz radio observations that indicate, from rotation measures, that radio emission from the western radio lobe in this source propagates over a shorter path length than radio emission from the eastern lobe \citep[e.g.,][]{carilli1997}. 

This configuration implies that the radio axis is also titled a non-zero inclination angle with respect to the projected plane as shown Fig. \ref{fig:absorption-cartoon}. If the jets were parallel to the plane of the sky, we should observe similar rotation measures at both the east and west radio lobes the rotation measure at the eastern lobe is higher than the one at the west. This jet-inclination can also explain the blueshifted emission being an outflow.


The kinematics surrounding the jet-gas interactions in the gas halo of Yggdrasil have been measured in previous studies. In \citet{villar-martin2003}, the narrow components of \ion{He}{II} emission are seen to be blueshifted from the systemic at velocity shifts of $\Delta \varv \sim$ 100 km s$^{-1}.$ This kinematically quiet component identified with a FWHM $\sim$ 500 km s$^{-1}$ is shown to extend across the entire halo, well beyond the radio hotspot, suggesting jet-gas interactions. In addition to this, \citet{humphrey2006} find that the \ion{He}{II} gas that has a perturbed component with FWHM = 1760 km s$^{-1}$ and the quiescent component with FWHM = 750 km s$^{-1},$ respectively, at a velocity shift of $\Delta \varv \sim$ -250 km s$^{-1}$ is low ionisation gas. In this work, we find that the blueshifted emission has FWHM = 970 km s$^{-1}$ at $\Delta \varv \sim$ -1000 km s$^{-1}.$ Although our measured velocity shifts differ slightly from those of previous works, the common thread is that the \ion{He}{II} emission is blueshifted relative to systemic and aligned with the radio axis indicating jet-gas interactions.  

This view is in good agreement with observations from G16 who see direct evidence of the ionisation cone projected along the radio jet axis in the form of extended \ion{He}{II}, \ion{C}{IV} and \ion{C}{III]} emission to the south-west of the source. An in-depth discussion of the ionised gas kinematics is presented in S18. Their interpretation differs from ours only slightly in that our radio axis is tilted relative to the projected plane. 

In addition to \ion{He}{II} enhancement, a dust continuum and enhanced star-formation rates have been observed to the south-west of the inferred centre of the galaxy halo. In G16, the strong dust continuum may be a result of star-bursts. This is supported by measured star-formation rates SFR = 41 M$_\odot$ yr$^{-1}$  in Yggdrasil compared to SFR = 747 M$_\odot$ yr$^{-1} $ in the dusty regions at the south-west or companion sources \citep{falkendal2019}. Anti-correlations between radio size and 1.1-mm luminosities (that trace SF) measured from a sample of 16 HzRGs in \citet{humphrey2011} have explained why more evolved sources, such as Yggdrasil, have experienced a decline in their star-formation. 

Furthermore, at z $>$ 3.0, metal enrichment and disturbed kinematics of extended gas haloes have been linked to jet-induced star-formation in HzRGs \citep[e.g.,][]{reuland2007}. One cannot say that there is jet-induced star-formation in this galaxy, however, because the projected radio size of $r \simeq 29$ kpc shows that the radio emission has not propagated far enough to induce star-formation at $r=65$ and 80 kpc from the host galaxy where the companion sources are located (G16).

\subsection{Possible origins and the nature of the absorbing gas}

We have shown that resonance line scattering or absorption of \ion{C}{IV}, \ion{N}{V} and \ion{Si}{IV} photons occurs at the same velocities as that of the strong \lya absorber (absorber 2). Due to the uncertainties in measuring \ion{Si}{IV} absorption, we will frame our discussion of the main absorber around Ly$\alpha,$ \ion{N}{V} and \ion{C}{IV} detections only. We obtain redshifts, column densities and Doppler parameters for absorbers in the \lya profile and find that the \ion{H}{I} column densities measured are consistent with previous results (\citealp{vanojik1997,jarvis2003,wilman2004}; G16; S18). 

The formation of absorber 2 is well supported by simulations in which the neutral gas column is formed at the bow shock of radio jets and cools to T $\simeq$ $10^4-10^6$ K before fragmenting due to Rayleigh-Taylor (RT) instabilities brought about by the radio cocoon \citep{krause2002,krause2005}. S18 also showed that absorber 2 may be outflowing which is in agreement with this framework. In addition to the high column density absorber, we also observe three \ion{H}{I} absorbers with column densities of $N_\ion{H}{I}/{\rm cm}^{-2} = 10^{13}-10^{14}.$ The weak absorbers (absorbers 1, 3 and 4) could be fragmented shells of gas that have been disrupted by RT instabilities, hence their low column densities. In this scenario, the absorbers are formed through the ageing of the radio source \citep{wilman2004,binette2006}. 

Absorber 2 has a much higher \ion{H}{I} column density of $N_\ion{H}{I}/{\rm cm}^{-2} = 10^{19.2}.$ It may be a metal-poor gas shell that was expelled from the galaxy by an early AGN-related feedback event \citep{jarvis2003,binette2006}. We agree with this interpretation given both the strength of the absorber and the sum of its ionised and neutral mass, $M_{\rm T}/M_\odot \geq 5.7\e{9},$ which is an order of magnitude lower than the galaxy's stellar mass component of $M_*/M_\odot = 1.2\e{11}$ implying that a very energetic event would have to be responsible for expelling such a vast amount of gas. 

Absorber 2, if photoionised primarily by the AGN has an excess nitrogen abundance a factor of 10 greater than its Solar abundance. Since absorber 2 has a low metallicity overall, it may be nitrogen enhanced as a result of stellar winds. Chemically enriched gas from the ISM may have been swept up by stellar winds and progressively diluted the outflowing gas shell. 

This scenario is similar to that of the z = 3.09 star-forming galaxy of \citet{wilman2005} which also contains an emission line region covered by a neutral gas shell of \ion{H}{I} column density, $N_\ion{H}{I}/{\rm cm}^2 \simeq 10^{19}$. The main distinction between this source and ours is the SFR which are much higher in the star-forming galaxy. If the mechanism by which the strong absorber in Yggdrasil has been enriched is a star-burst driven super-wind, it could mean that this process occurred at earlier epoch and we are now observing the galaxy after a major star-burst period has ceased. 

\section{Summary}\label{section:summary}


We have examined the absorbing gas surrounding the host galaxy of  MRC 0943-242, Yggdrasil, using MUSE data. Our results prove that the \ion{H}{I} absorbers measured from the \lya line are at the same velocities as \ion{C}{IV} and \ion{N}{V} absorbers. The new integral field unit (IFU) dataset shows that the high \ion{H}{I} column density absorbing gas (\lya absorber 2) with $N_\ion{H}{I}/{\rm cm}^{-2} = 10^{19.2}$  is a non-isotropic gas medium extending outwards from $r \gtrsim$ 60 kpc. It hydrogen fraction of X$_\ion{H}{I} \gtrsim 0.8$ which makes it a predominantly ionised cloud with a total estimated mass of the neutral and  ionised gas being, $M/M_\odot \geq 5.7\e{9}.$ Detections of \ion{Si}{II} absorption at the same velocity of this absorber suggest that it is ionisation bounded.

Similar to previous studies, we observe a diffuse extension of \ion{He}{II} gas which in alignment with the VLA detected radio axis and the {\it HST} detected optical/UV broad-band emission. The ionised gas is interpreted as a jet-induced outflow. The combination of radio axis and the observed \ion{He}{II} emission in this IFU data indicates that radio jet is tilted at an non-zero inclination angle. This finding is well supported by rotation measures in the east and west radio lobes of the galaxy which indicate that the emission from the eastern radio lobe has travelled further along the line-of-sight than that emitted from the western lobe.

We also estimate approximately the locations of the absorbers within the halo, assuming that they are outflowing. It is likely that absorber 2 is located at a greater distance from the AGN than absorber 1. Furthermore, the measured absorber  column densities in Yggdrasil are similar in magnitude to those of absorbers at velocity shifts of $\lvert \Delta \varv \rvert < 5000$ km s$^{-1}$ from quasars. 

Photoionisation models of the absorbing gas in \pkg{cloudy} has shown that ionising radiation from the AGN (for which we assume a spectral energy distribution with a power-law shape of $\alpha=-1$) is capable of producing the \ion{C}{IV} and \ion{N}{V} column densities observed in absorber 2 when the gas has a metallicity of Z/Z$_\odot = 0.01$ with nitrogen abundance enhanced by a factor of 10 relative to the Solar abundance (i.e., N/H = 10). This high column density absorber is interpreted as primordial gas that was propelled outward by an earlier feedback event (possibly AGN). The gas has subsequently been ionised by the AGN and chemically enriched by star-burst activity. 

In conclusion, this work shows the potential use of wide field integral field unit instruments in understanding  the configuration of complex systems such as the haloes of HzRGs. Similar future analysis will be fundamental in unveiling the different gas structures surrounding HzRGs, ultimately adding important constraints on the physics of these objects.

\section*{Acknowledgements}
SK acknowledges the International Max Planck Research Schools (IMPRS) and the European Southern Observatory (ESO). All authors acknowledge the ESO Paranal Observatory as the source for VLT/MUSE data that formed the basis of this work. SK, CDB and JV thank Richard Wilman and Matt J. Jarvis for the ancillary UVES spectrum. Many thanks to Paola Caselli and Ian Smail for providing helpful guidance and suggestions. 
 
This study has made use of data from the European Space Agency (ESA) mission
{\it Gaia} (\url{https://www.cosmos.esa.int/gaia}), processed by the {\it Gaia}
Data Processing and Analysis Consortium (DPAC,
\url{https://www.cosmos.esa.int/web/gaia/dpac/consortium}). Funding for the DPAC
has been provided by national institutions, in particular the institutions
participating in the {\it Gaia} Multilateral Agreement.

This work is on observations collected at the European Southern Observatory under ESO programmes 096.B-0752(A) and 068.B-0086(A). It is also based on observations made with the NASA/ESA Hubble Space Telescope, and obtained from the Hubble Legacy Archive, which is a collaboration between the Space Telescope Science Institute (STScI/NASA), the Space Telescope European Coordinating Facility (ST-ECF/ESA) and the Canadian Astronomy Data Centre (CADC/NRC/CSA).

MVM acknowledges support from the Spanish Ministerio de Ciencia, Innovación y Universidades (former Ministerio de Econom\'\i a y Competitividad) through the grant AYA2015-64346-C2-2-P. AH acknowledges FCT Fellowship SFRH/BPD/107919/2015; Support from
European Community Programme (FP7/2007-2013) under grant agreement
No. PIRSES-GA-2013-612701 (SELGIFS); Support from FCT through national funds (PTDC/FIS-AST/3214/2012 and UID/FIS/04434/2013), and by FEDER through COMPETE (FCOMP-01-0124-FEDER-029170) and COMPETE2020 
(POCI-01-0145-FEDER-007672). 

AH acknowledges support from the FCT-CAPES Transnational Cooperation Project "Parceria
Estrat\'egica em Astrof\'{I}sica Portugal-Brasil".

\bibliographystyle{aa}
\bibliography{0943_absorption}

\end{document}